\documentclass[12pt]{article}
\usepackage{graphicx}
\usepackage{epstopdf}
\usepackage{epsfig}
\usepackage{caption}
\usepackage{subcaption}
\usepackage{subfig}

\usepackage{epsfig,amsmath,amssymb,verbatim,mathrsfs}
\def\beq{\begin{eqnarray}}
\def\eeq{\end{eqnarray}}
\def\bea{\begin{eqnarray}}
\def\eea{\end{eqnarray}}

\newcommand{\lw}{\text{\tiny W}}
\newcommand{\lz}{\text{\tiny Z}}
\newcommand{\lf}{\text{\tiny F}}

\newcommand{\f}{\mathcal{F}}
\newcommand{\sqth}{\sqrt{3}}

\newcommand{\be}{\begin{equation}}
\newcommand{\ee}{\end{equation}}

\newcommand{\s}{\text{\tiny S}}

\begin{document}

\setlength{\baselineskip}{0.2in}

%\begin{comment}

%\twocolumn[\hsize\textwidth\columnwidth\hsize\csname
%@twocolumnfalse\endcsname
%%
%%
\begin{titlepage}
\noindent
%\flushright{October 2013}
%\begin{flushright}
%\end{flushright}
%%
%%
\vspace{0.2cm}

\begin{center}
  \begin{Large}
    \begin{bf}
Probing Exotic Fermions from a Seesaw/Radiative Model at the LHC

     \end{bf}
  \end{Large}
\end{center}

\vspace{0.2cm}

\begin{center}

\begin{large}
{Kristian~L.~McDonald}\\
     \end{large}
\vspace{0.5cm}
  \begin{it}
ARC Centre of Excellence for Particle Physics at the Terascale,\\
School of Physics, The University of Sydney, NSW 2006, Australia\\\vspace{0.5cm}
\vspace{0.1cm}
klmcd@physics.usyd.edu.au
\end{it}
\vspace{0.5cm}

\end{center}

%\center{\today}

\begin{abstract} 

There exist tree-level generalizations of the  Type-I and Type-III seesaw mechanisms that realize neutrino mass via low-energy effective operators with $d>5$. However, these generalizations also give radiative masses that can dominate the seesaw masses in regions of parameter space --- i.e.~they are not purely seesaw models, nor are they purely radiative models, but instead they are something in between.  A recent work detailed the  remaining minimal models of this type. Here we study the remaining  model with  $d=9$ and investigate the collider phenomenology of the exotic quadruplet fermions it predicts. These exotics can be pair produced at the LHC via electroweak interactions and their subsequent decays produce a host of multi-lepton signals. Furthermore, the branching fractions for events with distinct charged-leptons encode information about both the neutrino mass hierarchy and the leptonic mixing phases.   In large regions of parameter-space discovery at the LHC with a 5$\sigma$ significance is viable for masses approaching the TeV scale.
\end{abstract}

\vspace{1cm}

\end{titlepage}
%\pacs{PACS numbers: }
%]

%\setcounter{footnote}{1}
%\setcounter{figure}{0}
%\setcounter{table}{0}

%\tableofcontents

\vfill\eject

%\end{comment}

\tableofcontents
\pagenumbering{gobble}
\newpage
%%%%%%%%%%%%%%%%%%%%%%%%%%%%%%%%%%%%%%%%%%%%%%%%%%%%%
\section{Introduction\label{sec:introduction}}
\pagenumbering{arabic}
The Type-I~\cite{type1_seesaw} and Type-III~\cite{Foot:1988aq} seesaw mechanisms offer a simple explanation for the existence of light Standard Model (SM) neutrinos. In these approaches the tree-level exchange of heavy intermediate fermions achieves neutrino masses with an inverse dependence on the heavy-fermion mass, $m_\nu\sim \langle H^0\rangle^2/M_\f$, suppressing the masses relative to the weak scale.   In the low-energy effective theory these  masses are described by the non-renormalizable operator $\mathcal{O}_\nu=(LH)^2/\Lambda$,  which famously has   mass-dimension $d=5$~\cite{Weinberg:1979sa}.

\begin{figure}[h]
\begin{center}
        \includegraphics[width = 0.4\textwidth]{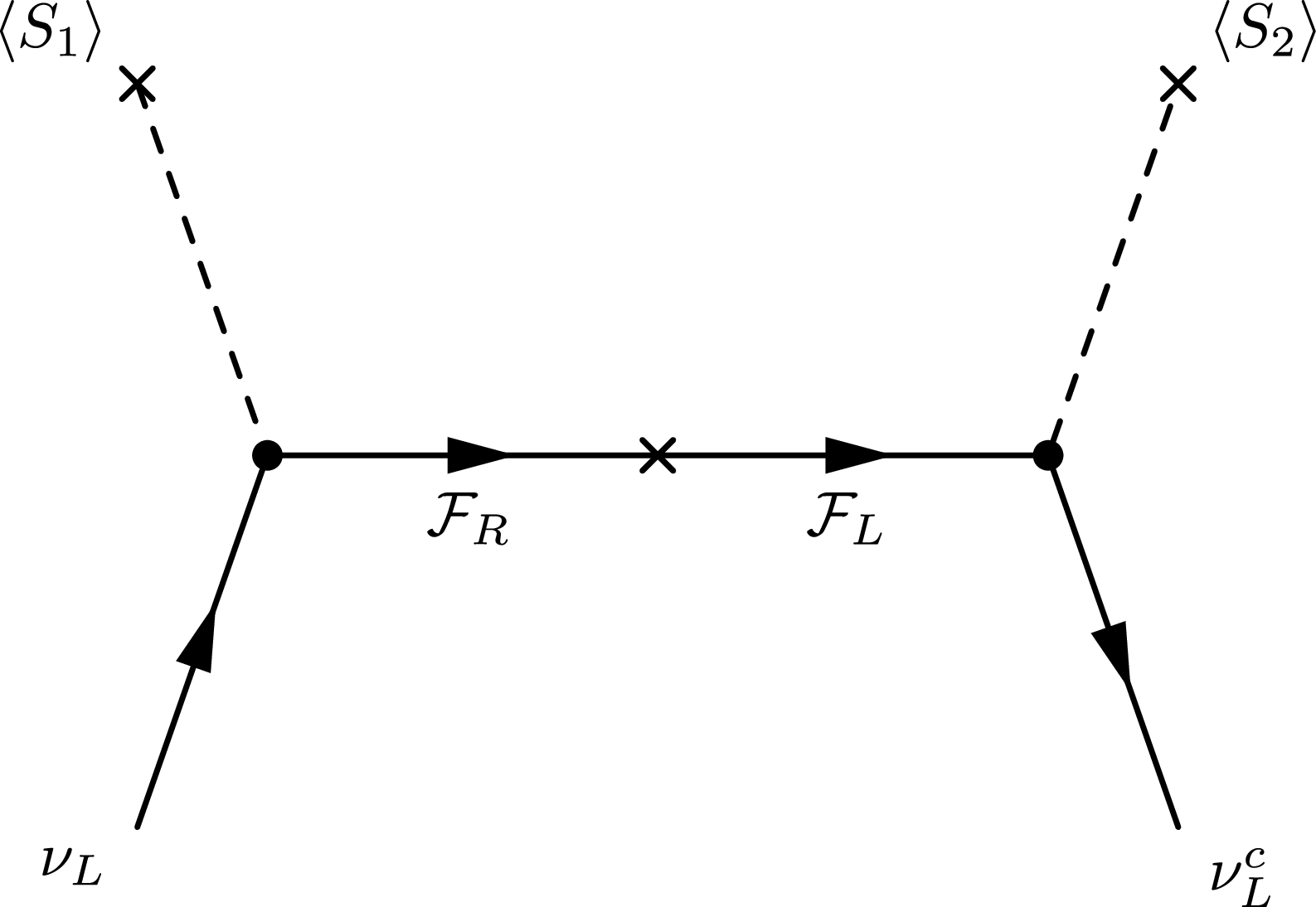}
\end{center}
\caption{Generic tree-level diagram for a seesaw mechanism with a heavy intermediate fermion.  The simplest realizations are the Type-I and Type-III seesaws, for which the external scalars are the SM doublet, $S_1=S_2=H\sim(1,2,1)$, and $\f_L\equiv \f_R^c$ is a Majorana fermion. In the generalized seesaws the fermion can be either Majorana or Dirac and the external scalars can be beyond-SM multiplets.}\label{fig:L_vertex_nu_tree_generic}
\end{figure}

There exist generalizations of the Type-I and Type-III seesaws that can similarly explain the existence of light SM neutrinos. The basic point is that the Type-I and Type-III seesaws can be described by a generic tree-level diagram with two external scalars and a heavy intermediate fermion; see Figure~\ref{fig:L_vertex_nu_tree_generic}. The use of different intermediate fermions allows for variant tree-level seesaws, where either one or both of the external scalars is a beyond-SM field.

Naively it appears that many variant seesaws are possible. However, the vacuum expectation values (VEVs) of the beyond-SM scalars are generally constrained by $\rho$-parameter measurements to satisfy $\langle S_{1,2}\rangle\lesssim\mathcal{O}(\mathrm{GeV})$. Such small VEVs can arise naturally if they are induced and therefore develop an inverse dependence on the scalar masses, i.e.~$\langle S_{1,2}\rangle\propto M_{1,2}^{-2}$.  Demanding such an explanation for the small VEVs greatly restricts the number of minimal realizations of Figure~\ref{fig:L_vertex_nu_tree_generic}~\cite{McDonald:2013kca}.  Because the VEVs of the beyond-SM scalars are induced, these generalized seesaws generate low-energy effective operators with $d>5$. It was  shown that there are only four such minimal models that give  effective operators with $d\le 9$~\cite{McDonald:2013kca}; namely the $d=7$ model of Ref.~\cite{Babu:2009aq}, the $d=9$ models of Refs.~\cite{Kumericki:2012bh,Liao:2010ku} and Ref.~\cite{Picek:2009is}, and the $d=9$ model proposed in Ref.~\cite{McDonald:2013kca}.\footnote{We list the particle content for these models, and show the explicit $d>5$ nature of the associated Feynman diagrams, in Section~\ref{sec:new_seesaw}; see Table~\ref{table:seesaw_models}  and Figure~\ref{fig:general_seesaw_diagrams} respectively.}

%---------------------------------------------------------
%---------------------------------------------------------

\begin{figure}[t]
\begin{center}
        \includegraphics[width = 0.5\textwidth]{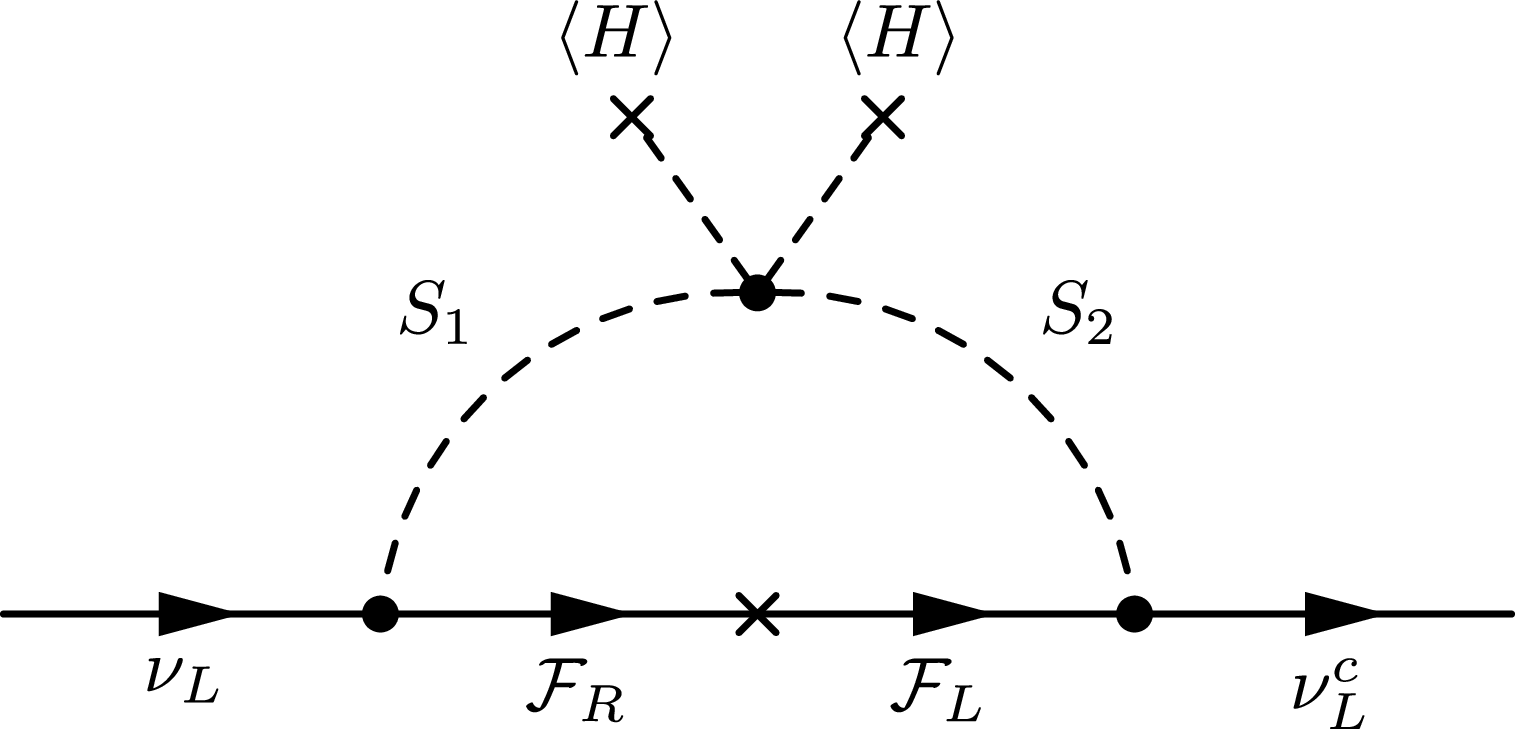}
\end{center}
\caption{The generic loop-diagram present in any model that realizes the tree-level seesaw in Figure~\ref{fig:L_vertex_nu_tree_generic}.}\label{fig:loop}
\end{figure}

\begin{figure}[ttt]
\begin{center}
        \includegraphics[width = 0.6\textwidth]{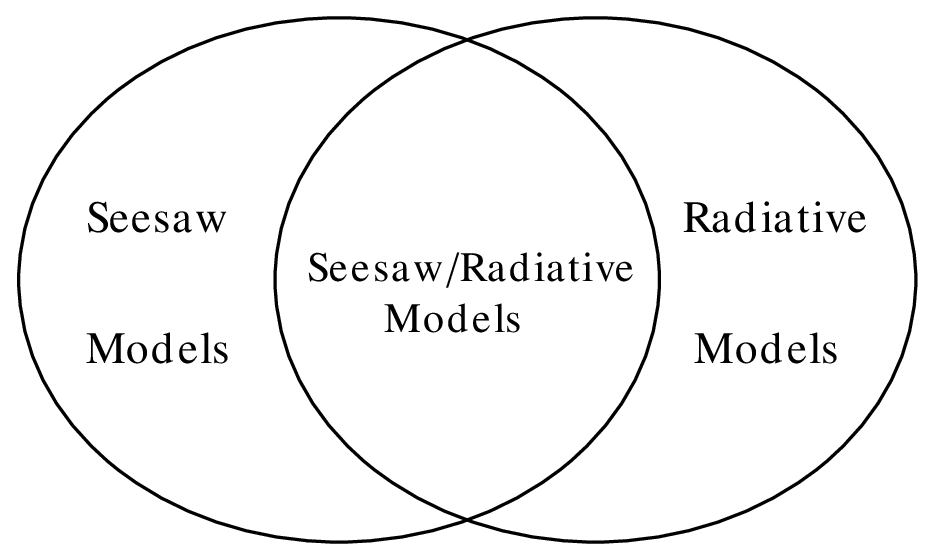}
\end{center}
\caption{A Venn diagram for a portion of theory space with massive neutrinos.}\label{fig:2venn}
\end{figure}

The generalized seesaws  turn out to be more complicated creatures than their Type-I and Type-III cousins.  Extensions of the SM that permit the tree-level diagram of Figure~\ref{fig:L_vertex_nu_tree_generic} automatically admit the term $\lambda H^2S_1S_2\subset V(H,S_1,S_2)$ in the scalar potential~\cite{McDonald:2013kca}. This allows one to close the seesaw diagram in Figure~\ref{fig:L_vertex_nu_tree_generic} to obtain the $d=5$ loop-diagram in Figure~\ref{fig:loop}. Thus, strictly speaking,  the generalized models  are not purely seesaw models, nor are they purely radiative models, but instead they are something in between. Both mechanisms are always present, with the tree-level mass being dominant in some regions of parameter space and the radiative mass being dominant in other regions. If one envisions the theory space for models with massive neutrinos, the generalized seesaws exist in the intersection of the set of models with seesaw masses and the set of models with radiative masses  (see Figure~\ref{fig:2venn}).    Put succinctly, these are seesaw/radiative models, and they are irreducible, in the sense that modifying the particle content to remove one effect necessarily removes them both. As we shall see, this is manifest in an identical flavor structure for the seesaw and radiative masses.

%---------------------------------------------------------
%---------------------------------------------------------

These distinctions give an important difference relative to the Type-I and Type-III approaches. In the seesaw/radiative models neutrino mass is generated either by a tree-level seesaw described  by the operator $\mathcal{O}_\nu=L^2 H^{d-3}/\Lambda^{d-4}$ with $d>5$, or by a radiative diagram generating a $d=5$ operator with additional loop-suppression. In either case the new physics is constrained to be much lighter than that allowed by the Type-I and Type-III seesaws; e.g.~one can have $\Lambda\lesssim (\langle H\rangle^{d-3}/m_\nu)^{\frac{1}{d-4}}$ which decreases with increasing $d$. Collider experiments at the energy frontier will therefore explore the parameter space for these generalized models long before the full parameter space for the Type-I and Type-III seesaws can be investigated.

In the present work, we detail the nature of neutrino mass in the newly proposed model with $d=9$, and  study the collider phenomenology of the exotic fermions predicted by the model.  These fermions form an isospin-3/2 representation of $SU(2)_L$ and contain a doubly-charged component.  Collider production of the exotics is controlled by electroweak interactions and  depends only on the fermion mass.  The decay properties of the fermions, and therefore the expected signals at colliders like the LHC, have some sensitivity to model details and, in particular, depend on the mixing with SM leptons. However, because the model predicts a basic relation amongst VEVs ($\langle S_1\rangle \gg\langle S_2\rangle$), some decay branching fractions can be largely determined; e.g.~the total leptonic branching fractions can be determined with essentially no dependence on the neutrino mass hierarchy. However, the relative branching fractions for decays to \emph{different} charged leptons have remnant dependence on the properties of the neutrino sector. We shall see that the number of light charged-lepton events ($\ell=e,\,\mu$), relative to the number of tauon events, can encode information regarding  the neutrino mass hierarchy and the mixing phases. 

%---------------------------------------------------------
%---------------------------------------------------------

A number  of  signals predicted by the model are reminiscent of those found in  related seesaw models like the Type-III seesaw and the $d=9$ models~\cite{Kumericki:2012bh,Picek:2009is}. However, the branching fractions for lepton-number violating like-sign dilepton events is suppressed relative to that found in the Type-III seesaw~\cite{delAguila:2008cj}. In our analysis we discuss  differences 
between the models and indicate strategies for searching for the exotic fermions; for example, the present model predicts an unobservable rate for lepton-number violating events like $\ell^\pm\ell^\pm W^\mp W^\mp$, whereas events like  $\ell^\pm\ell^\pm W^\mp  Z$ are  expected in both the Type-III case~\cite{delAguila:2008cj} and the $d=9$ model of Ref.~\cite{Kumericki:2012bh}. Such events, and others that we outline, lead to a host of multi-lepton final states.  We shall see that the model also predicts a doubly-charged fermion that can be discovered at the 5$\sigma$ level for masses approaching the TeV scale in optimistic cases.

In our presentation we make efforts to follow the structure of Refs.~\cite{Kumericki:2012bh} and~\cite{Picek:2009is}, which detail the collider phenomenology of related exotic fermions, to allow for easier comparison. As shall be evident during our analysis, there are aspects of our work that are relevant for the related models.  Our results suggest it would be interesting to undertake a detailed comparative analysis  of the exotic fermions appearing in the models with $d\le9$, including the triplets $\f\sim(1,3,2)$, which appear in the $d=7$ model~\cite{Babu:2009aq},\footnote{To date the collider phenomenology of these states have not been studied within the context of the $d=7$ model. A study of the bounds from flavor changing processes appeared in Ref.~\cite{Liao:2010rx}, and they were studied in different contexts in Ref.~\cite{DelNobile:2009st}.} and the quintuplet fermions from the alternative $d=9$ models~\cite{Kumericki:2012bh,Picek:2009is}.  All four models with $d\le9$ contain doubly-charged exotic fermions; searches for these states  would appear to provide a simple way to obtain generic experimental constraints for this class of models.

Many works have studied  production mechanisms and detection prospects for the heavy neutrinos employed in the Type-I seesaw~\cite{Keung:1983uu}.  Both CMS~\cite{Chatrchyan:2012fla} and ATLAS~\cite{ATLAS:2012yoa} have  searched for the corresponding  signals in the LHC data. Similarly,  the  triplet fermions in the Type-III seesaw  are well studied~\cite{Franceschini:2008pz}, and Ref.~\cite{ATLAS:2013hma} (\cite{CMS:2012ra}) contains an ATLAS (CMS) search for these exotics. A  comparative study of LHC signals from the $d=5$ seesaws has appeared~\cite{delAguila:2008cj}, and more general discussion of TeV-scale exotics related to neutrino mass~\cite{Chen:2011de}, and right-handed neutrinos~\cite{Drewes:2013gca}, exists in the literature. For the collider phenomenology of the exotic scalars in the $d=7$ model see Ref.~\cite{Bambhaniya:2013yca} (also see~\cite{Ren:2011mh}). Note that perturbative unitarity gives general upper-bounds on the quantum numbers of larger multiplets~\cite{Hally:2012pu,Earl:2013jsa}, and that the quadruplet fermions of interest in this work were previously considered as dark matter~\cite{Cirelli:2005uq}, and in relation to radiative neutrino-mass~\cite{Law:2013saa}.  Alternative models of neutrino mass realizing low-energy effective operators with $d>5$ exist~\cite{Bonnet:2009ej}, and an earlier work combined a seesaw model with a radiative model~\cite{Kubo:2006rm} (also see Ref.~\cite{Kajiyama:2013sza}).\footnote{The models of interest here differ from these earlier works. In Refs.~\cite{Kubo:2006rm,Kajiyama:2013sza} distinct beyond-SM fields generate the tree- and loop-masses; one can modify the particle spectrum to turn off one effect while retaining the other. In  the present models the same fields generate both the tree- and loop-masses, and experimentally viable masses can be achieved from either effect.} Also, it was recently shown that some versions of the inverse seesaw mechanism can generate neutrino mass via a combination of both tree-level and radiative masses, similar to the models discussed here~\cite{Dev:2012sg}.

The layout of this paper is as follows. In Section~\ref{sec:new_seesaw} we introduce the $d=9$ model and discuss the symmetry-breaking sector. Section~\ref{sec:neutral_fermion_mass} details the origin of neutrino mass and Section~\ref{sec:applying_data}  investigates the extent to which the parameters can be fixed by neutrino oscillation data.  Collider production of exotic fermions is discussed in Section~\ref{sec:production_of_fermions}. The mass-eigenstate interaction Lagrangian is presented in Section~\ref{sec:mass_eigenstate_ints} and exotic fermion decays are detailed in Section~\ref{sec:exotic_fermion_decay}. Detection signals are discussed in Section~\ref{sec:signals} and the work concludes in Section~\ref{sec:conc}. 
%%%%%%%%%%%%%%%%%%%%%%%%%%%%%%%%%%%%%%%%%%%%%%%%%%%%%

%%%%%%%%%%%%%%%%%%%%%%%%%%%%%%%%%%%%%%%%%%%

%%%%%%%%%%%%%%%%%%%%%%%%%%%%%%%%%%%%%%%%%%%
\section{A  Seesaw/Radiative Model with $\mathbf{d=9}$\label{sec:new_seesaw}}
\begin{figure}[ttt]
\centering
\begin{tabular}{ccc}
\epsfig{file=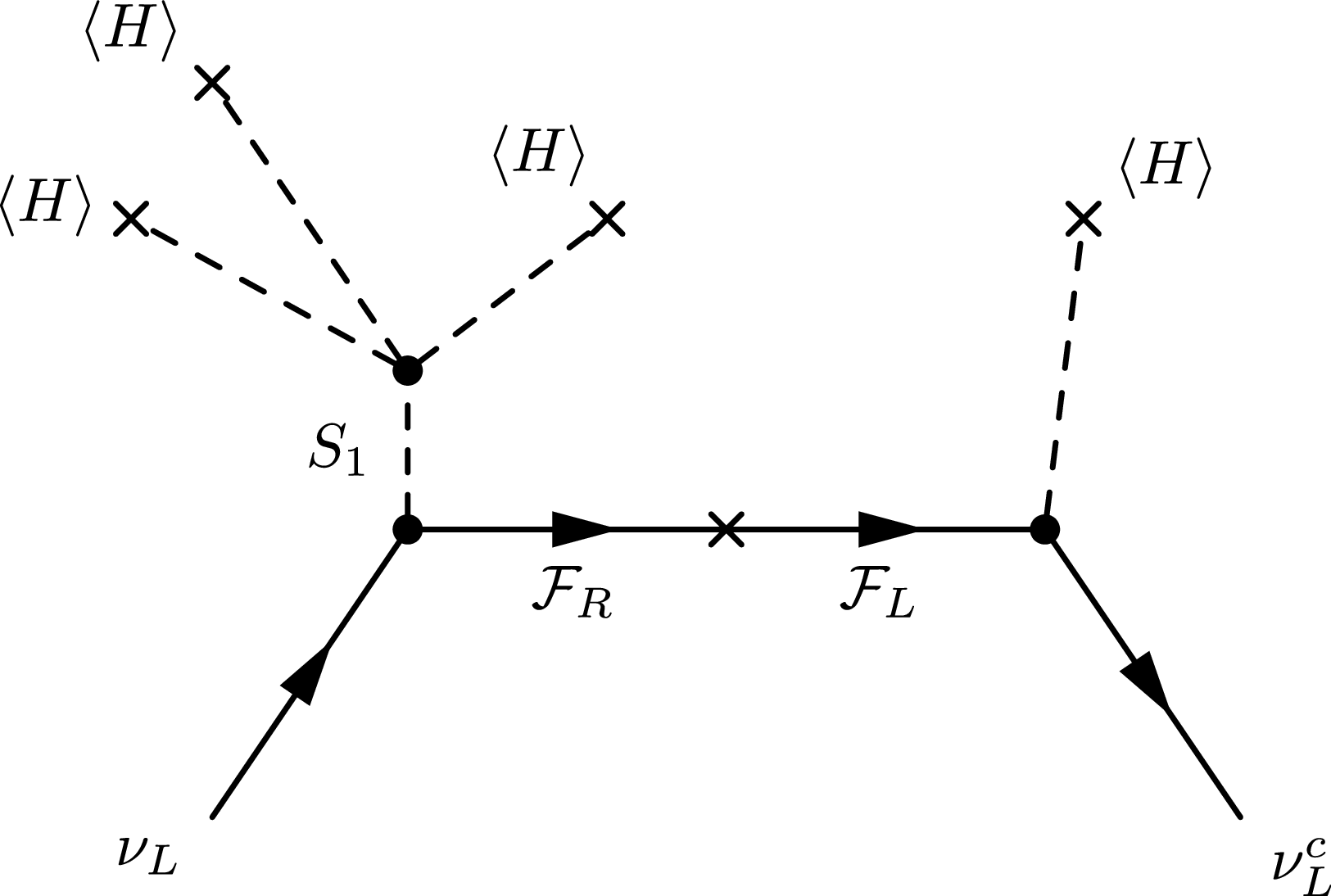,width=0.42\linewidth,clip=}&&
\epsfig{file=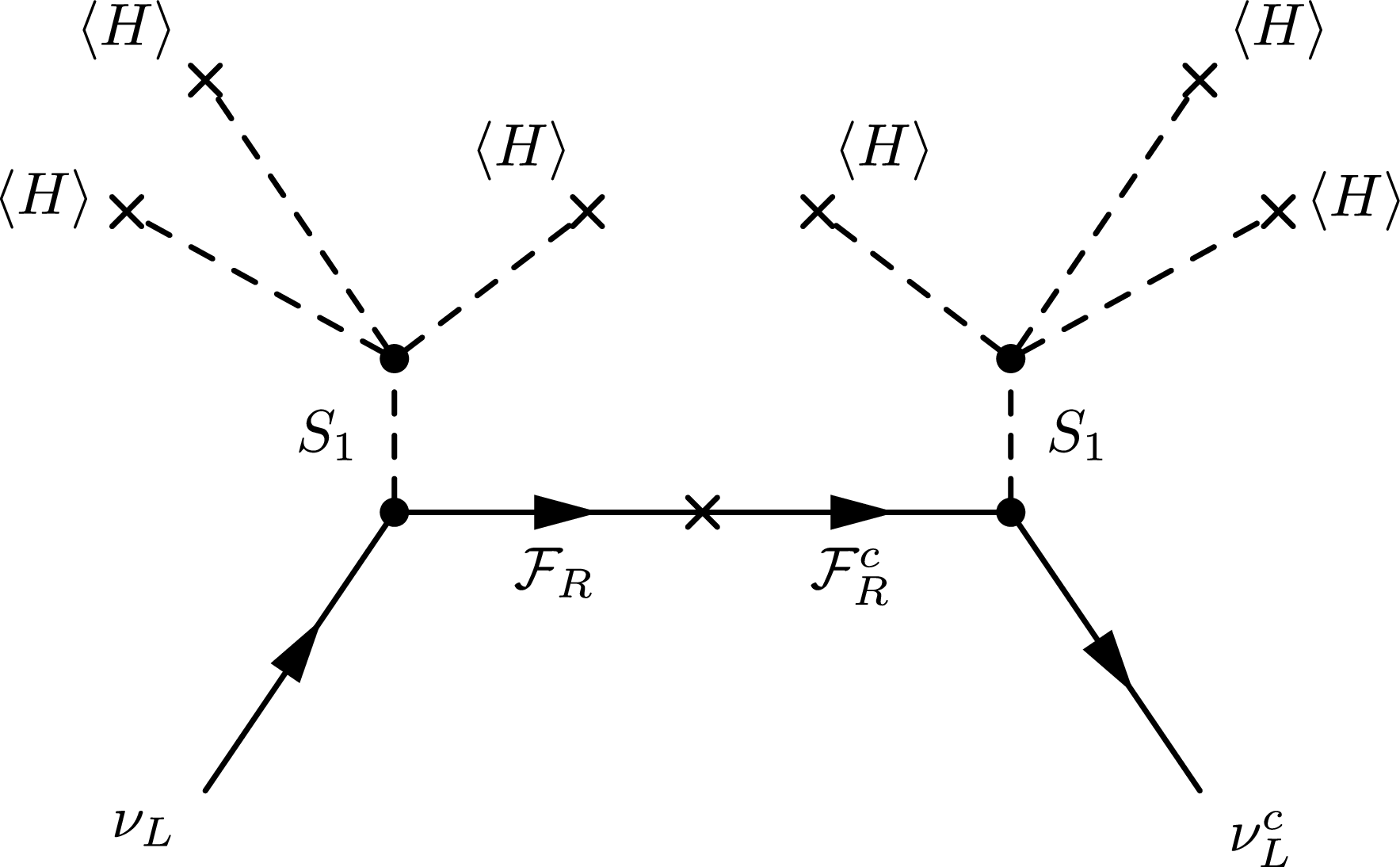,width=0.42\linewidth,clip=}\\
(a)&&(b)\\
&&\\
&&\\
\epsfig{file=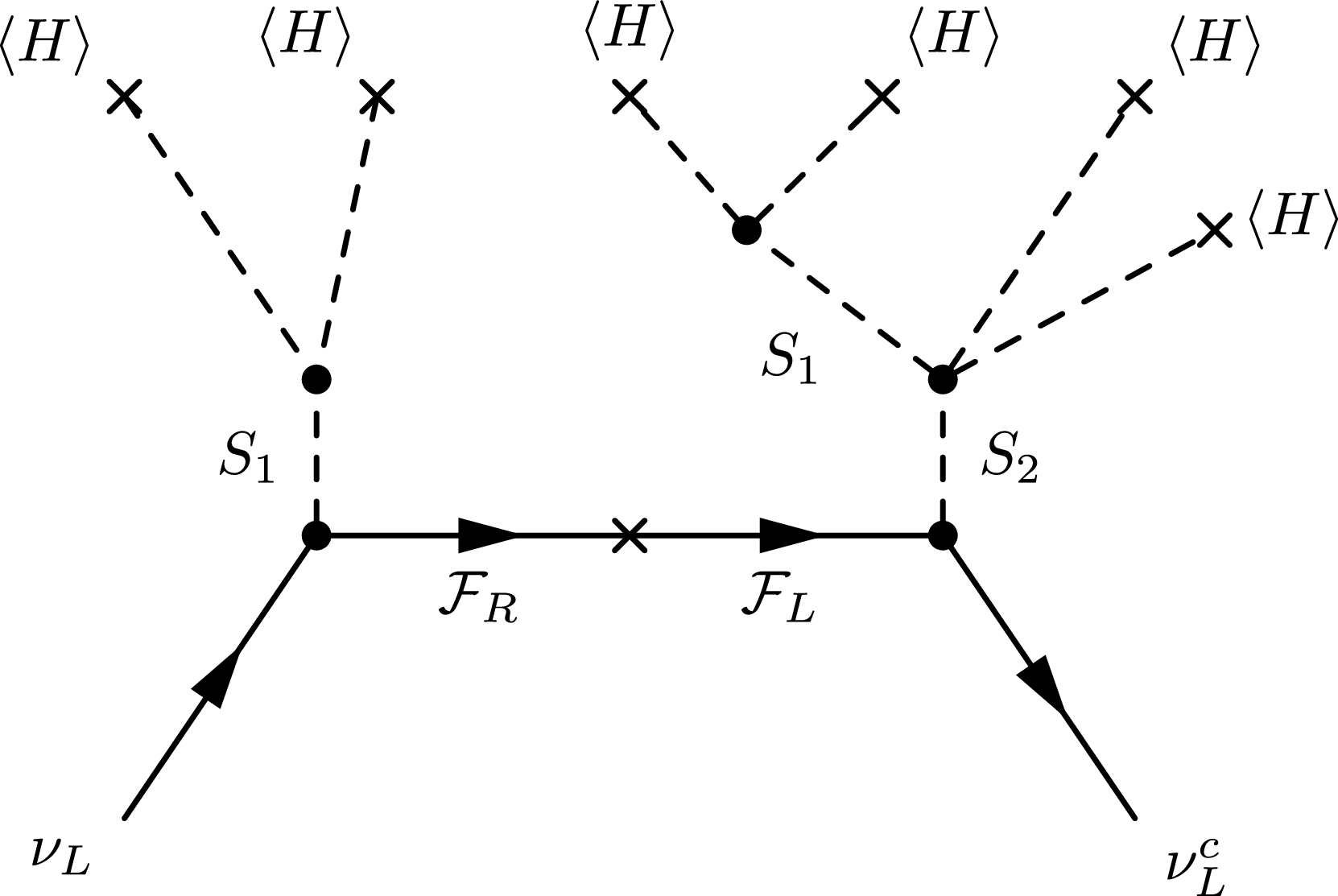,width=0.42\linewidth,clip=}&&
\epsfig{file=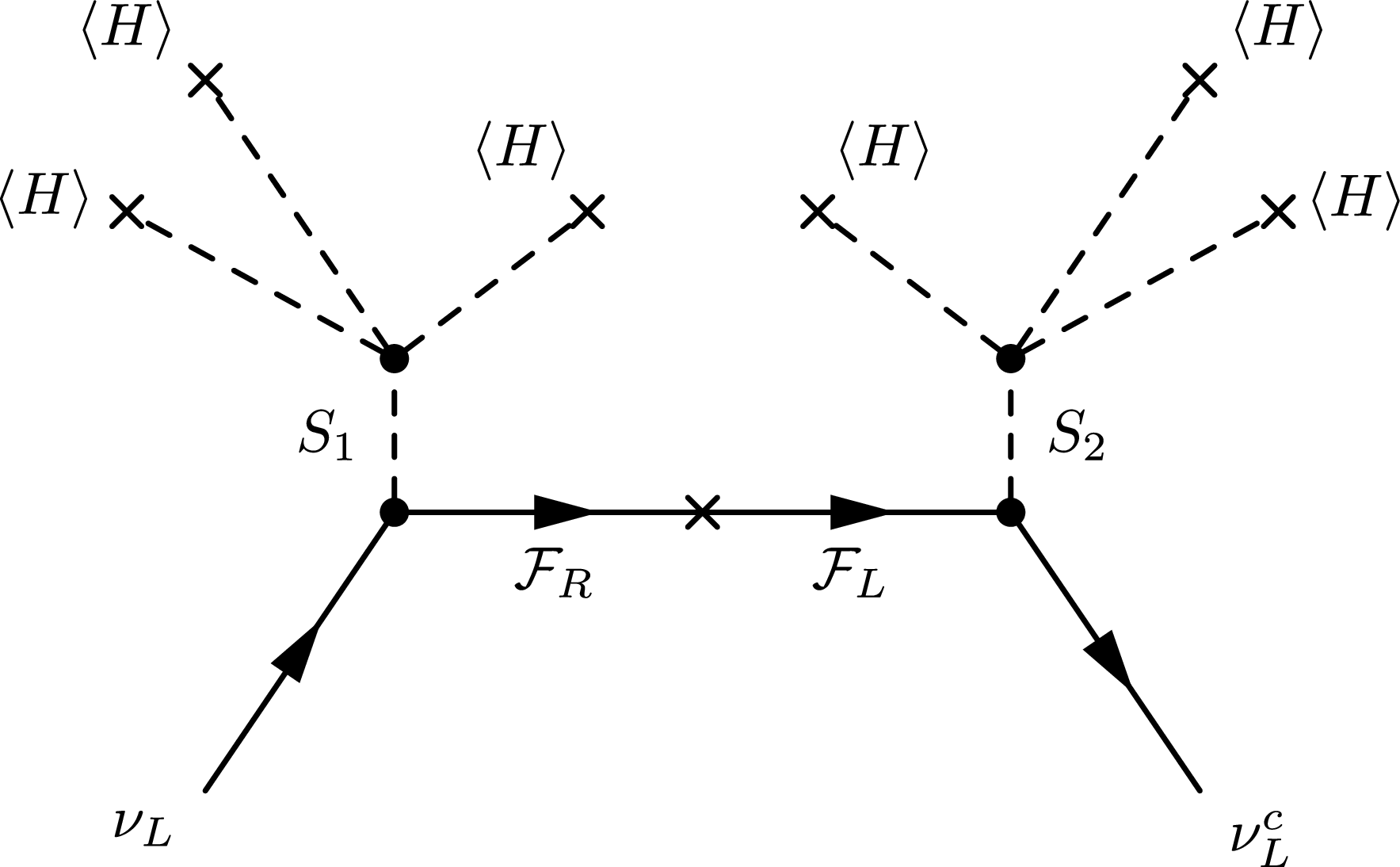,width=0.42\linewidth,clip=}\\
(c)&&(d)\\
&&\\
\end{tabular}
\caption{Feynman diagrams for the tree-level seesaws with $5<d\le9$.  Model $(c)$ is the focus of this work (labels match Table~\ref{table:seesaw_models}).}\label{fig:general_seesaw_diagrams}
\end{figure}

In this section we introduce the model of interest in this work and discuss aspects its scalar sector. As noted already, there are only four minimal models of this type that produce low-energy effective operators with $5<d\le9$~\cite{McDonald:2013kca}. We list the particle content for these models in Table~\ref{table:seesaw_models}.\footnote{Note that model $(b)$ can also be implemented as an inverse seesaw mechanism~\cite{Law:2013gma}.} Each model generates the tree-level diagram in Figure~\ref{fig:L_vertex_nu_tree_generic} and can thus achieve seesaw neutrino masses. However,  because the VEVs of the beyond-SM scalars are induced the corresponding Feynman diagrams can be ``opened up." These ``open" Feynman diagrams reveal the $d>5$ nature of the associated low-energy operators and are shown in Figure~\ref{fig:general_seesaw_diagrams}. 

 The new $d=9$ model appears as model $(c)$ in the table and is obtained by adding the following multiplets to the SM:
\bea
\f&\equiv&\f_L+\f_R\ \sim\ (1,4,-1),\nonumber\\
S_1&\sim&(1,3,0)\ \equiv\ \Delta,\nonumber\\
S_2^*&\sim& (1,5, 2)\ \equiv\ S.
\eea
These permit the following pertinent Yukawa terms
\bea
\mathcal{L}\ \supset\ -\lambda_\Delta\,\bar{L} \f_R\Delta-\lambda_\Delta^\dagger \,\overline{\f_R}L\Delta -\lambda_{\s}\,\overline{L^c} \f_LS-\lambda_{\s}^\dagger\, \overline{\f_L}L^cS^*,\label{eq:new_yukawa}
\eea
whose explicit $SU(2)$ structure is 
\bea
\mathcal{L}&\supset&-\lambda_\Delta\,\bar{L}^a\,(\f_R)_{abc}\,\epsilon^{cd}\,\Delta_d^{\ b}-\lambda_\Delta^\dagger\, (\overline{\f_R})^{abc}\,L_a\,\Delta_b^{\ d}\,\epsilon_{cd}\nonumber\\
& &-\lambda_{\s}\,(\overline{L^c})^a\,S_{abcd}(\f_L)_{b'c'd'}\,\epsilon^{bb'}\,\epsilon^{cc'}\,\epsilon^{dd'}-\lambda_{\s}^*\, (\overline{\f_R})^{a'b'c'}\,(L^c)_d\,(S^*)^{abcd}\,\epsilon_{aa'}\,\epsilon_{bb'}\,\epsilon_{cc'}.
\eea
Here we write the fermion as a symmetric tensor $\f_{abc}$, with components\footnote{Note that $\f^-$ is not the anti-particle of $\mathcal{F}^+$: $\overline{\f^+}\ne\f^-$.}
\bea
\f_{111}= \f^+\,,\quad \f_{112}=\frac{1}{\sqrt{3}}\,\f^0\,,\quad \f_{122}\ \frac{1}{\sqth}\,\f^-\,,\quad\f_{222}=\f^{--}\,,
\eea
and similarly the scalar $S\sim(1,5,2)$ is represented by the symmetric tensor $S_{abcd}$, with components
\bea
S_{1111}=S^{+++}\,,\quad S_{1112}=\frac{1}{\sqrt{4}}S^{++}\,,\quad S_{1122}=\frac{1}{\sqrt{6}}S^+\,,\quad S_{1222}=\frac{1}{\sqrt{4}}S^0\,,\quad S_{2222}=S^-,
\eea
where one should differentiate between  $S^-$ and $(S^+)^*$.  The matrix form of the real triplet is taken as
\bea
\Delta&\equiv& \Delta_a^{\ b}\ =\ \frac{1}{2}\left(
\begin{array}{cc}
\Delta^0&\sqrt{2}\Delta^+\\
\sqrt{2}\Delta^-&-\Delta^0
\end{array}
\right).
\eea
With the above, one can expand the Yukawa couplings to obtain the Lagrangian terms of interest for the seesaw and radiative diagrams. The explicit expansions appear in Appendix~\ref{app:expanded_yukawa}. Note that flavor labels are suppressed in the above, so that $\lambda_\s=\lambda_{\s,\ell}$ with $\ell\in\{e,\,\mu,\,\tau\}$ etc.

%%%%%%%%%%%%%%%%%%%%%%%%%%%%%%%%%%%%%%%%%%%%%%%%%%%%%%%

\begin{table}
\centering
\begin{tabular}{|c|c|c|c|c|c|c|}\hline
& & & & &&\\
\ \ Model\ \ &
$S_1$ & $\mathcal{F}$ &
$S_2$& Mass Insertion&$\ \ [\mathcal{O}_\nu]\ \ $&Ref.\\
& & & & &&\\
\hline
%-----------------------------------------------%
& & & & &&\\
$(a)$& $(1,4,-3)$ &$(1,3,2)$ &$(1,2,1)$&Dirac&$d=7$ &\cite{Babu:2009aq} \\ 
& & & & &&\\
\hline
%------------------------------------------------%
& & & & &&\\
$(b)$& $(1,4,1)$ &$(1,5,0)$ &$-$&Majorana&$d=9$ &\cite{Kumericki:2012bh}\\ 
& & & & &&\\
\hline
%------------------------------------------------%
& & & & &&\\
$(c)$& $(1,3,0)$ &$(1,4,-1)$ &$(1,5,-2)$&Dirac&$d=9$ &\cite{McDonald:2013kca}\\ 
& & & & &&\\
\hline
%------------------------------------------------%
& & & & &&\\
$(d)$& $(1,4,-3)$ &$(1,5,2)$ &$(1,4,1)$&Dirac&$d=9$ &\cite{Picek:2009is}\\ 
& & & & &&\\
\hline
%------------------------------------------------%
\end{tabular}
\caption{\label{table:seesaw_models} Minimal Seesaw/Radiative Models with $d\le9$~\cite{McDonald:2013kca}.}
\end{table}

%%%%%%%%%%%%%%%%%%%%%%%%%%%%%%%%%%%%%%%%%%%%%%%%%%%%%

The scalar potential contains the terms
\bea
V(H,\Delta,S)&\supset& M_\Delta^2 \mathrm{Tr}[\Delta\Delta] +\mu H^\dagger \Delta H \label{eq:delta_potential_vev_terms}
\eea
where $H\equiv H_a=(H^+,\,H^0)^T$ is the SM doublet. The last term induces a VEV for $\Delta^0$ after electroweak symmetry breaking:
\bea
\langle \Delta^0\rangle &\simeq& \frac{\mu\langle H^0\rangle^2}{2M_\Delta^2}.\label{delta_vev}
\eea
The inverse mass-dependence in this expression shows that $\langle \Delta^0\rangle$ is naturally suppressed relative to the electroweak scale for $M_\Delta\gg \langle H^0\rangle\simeq 174$~GeV.
Similarly the terms\footnote{The expansion of the $\lambda$-term appears in  Appendix~\ref{app:expanded_yukawa}.}
\bea
V(H,\Delta,S)&\supset& M_\s^2S^* S- \lambda \tilde{H}^\dagger \Delta S^* H +\mathrm{H.c.},
\eea
trigger a nonzero VEV for $S^0$,
\bea
\langle S^0\rangle &\simeq& \lambda\frac{\langle \Delta^0\rangle \langle H^0\rangle^2}{2M_\s^2}\ \simeq\ \lambda\,\frac{\mu\,\langle H^0\rangle^4}{4M_\s^2\,M_\Delta^2}.\label{s_vev}
\eea
These expressions show that  $\langle S^0\rangle/\langle \Delta^0\rangle\simeq\lambda\langle H^0\rangle^2/(2M_\s^2)\ll1$ is generically expected for $\lambda\lesssim0.1$, given that direct searches for exotic charged fields require $M_\s\gtrsim\mathcal{O}(100)$~GeV. We work with $\lambda\lesssim0.1$ throughout so that $\langle S^0\rangle/\langle \Delta^0\rangle\ll1$. This is a rather generic feature of the model;  we shall see that it influences both the decay properties and collider signals of the exotic fermions. Note that in Eq.~\eqref{delta_vev} one can consider $M_{\Delta}$ to denote the full tree-level mass for $\Delta^0$,  containing both the explicit mass-term for $\Delta^0$ in Eq.~\eqref{eq:delta_potential_vev_terms}  and the additional subdominant contributions from the VEVs of the other scalars. Similarly for $M_\s$ in Eq.~\eqref{s_vev}. 

%---------------------------------------------------------
%----------------------------------------------------

\begin{figure}[ttt]
\begin{center}
        \includegraphics[width = 0.5\textwidth]{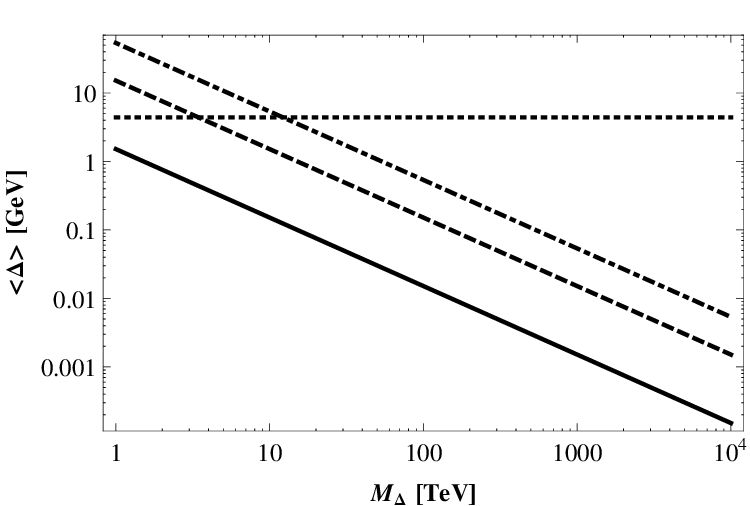}
\end{center}
\caption{The VEV for the scalar $\Delta\sim(1,3,0)$ as a function of the scalar mass $M_{\Delta}$, for fixed values of the dimensionful coupling $\mu$. The solid (dashed, dot-dashed) line is for $\mu/M_{\Delta}=0.1$ ($1,\,\sqrt{4\pi}$), and the horizontal line is the upper bound on $\langle \Delta\rangle$ from the $\rho$ parameter constraint.}\label{fig:delta_mass_v_vev}
\end{figure}

The beyond-SM scalars $S$ and $\Delta$ contribute to electroweak symmetry breaking and thus modify the tree-level value of the $\rho$ parameter. The SM predicts the tree-level value $\rho=1$~\cite{Ross:1975fq}, and the experimentally observed value is $\rho=1.0004^{+0.0009}_{-0.0012}$ at the $3\sigma$ level~\cite{Beringer:1900zz}. Consequently beyond-SM scalars with isospin $I_s\ne1/2$ must have small VEVs. In the present model the tree-level $\rho$ parameter is given by
\bea
\rho&\simeq& 1+2\frac{\langle \Delta^0\rangle^2}{\langle H^0\rangle^2}+6\frac{\langle S^0\rangle^2}{\langle H^0\rangle^2}.
\eea
The constraint requires
\bea
\langle \Delta^0\rangle^2 + 3 \langle S^0\rangle^2\ \lesssim\ 20~\mathrm{GeV}^2\,,
\eea
which reduces to
\bea
\langle \Delta^0\rangle&\lesssim &4.4~\mathrm{GeV}\qquad\mathrm{for}\quad \langle  S^0\rangle\ll\langle\Delta^0\rangle.
\eea
Thus, we generically require $\langle  S^0\rangle,\,\langle\Delta^0\rangle\lesssim1$~GeV. Such small VEVs arise naturally due to the inverse dependence on the scalar masses found in Eqs.~\eqref{delta_vev} and \eqref{s_vev}. We plot the VEV $\langle\Delta^0\rangle$ as a function of $M_\Delta$ for the fixed values of $\mu/M_{\Delta}=\{0.1,\,1,\,\sqrt{4\pi}\}$ in Figure~\ref{fig:delta_mass_v_vev}.

%---------------------------------------------------------
%---------------------------------------------------------

The nonzero VEVs for $\Delta^0$ and $S^0$ induce mixing between these scalars and the SM Higgs. For  $\langle S^0\rangle/\langle \Delta^0\rangle\ll1$ the mixing of $S^0$ with the Higgs is subdominant to the $\Delta^0$-$ H^0$ mixing, and to good approximation one can neglect the $S^0$-$H^0$ mixing. Shifting the neutral scalars around their VEVs, $H^0\rightarrow\langle H^0\rangle + (h^0 +i\eta^0)/\sqrt{2}$ and $\Delta^0\rightarrow \langle\Delta^0\rangle+\Delta^0$, the results of Ref.~\cite{FileviezPerez:2008bj} allow one to approximate the neutral-scalar mixing as
\bea
\left(
\begin{array}{c}
h_1\\h_2
\end{array}
\right)=
\left(
\begin{array}{cc}
\cos\theta_0&\sin\theta_0\\
-\sin\theta_0&\cos\theta_0
\end{array}
\right)
\left(
\begin{array}{c}
h^0\\\Delta^0
\end{array}
\right),
\eea
where $h_{1,2}$ are the mass eigenstates. For $\mu\gtrsim \langle H^0\rangle$ the mixing angle obeys 
\bea
\tan2\theta_0\ \approx\ \frac{2\sqrt{2}\langle \Delta^0\rangle}{\langle H^0\rangle}\ \ll\ 1,
\eea
 giving $\theta_0\approx \sqrt{2}\langle \Delta^0\rangle/\langle H^0\rangle\ll1$. In what follows we denote the mostly-SM Higgs $h_1$, with mass $m_1\simeq125$~GeV, simply as $h$. Unfortunately the tiny mixing angle $\theta_0$ will not be discernible at the LHC for the parameter space of interest in this work (see Ref.~\cite{FileviezPerez:2008bj} for details).
%%%%%%%%%%%%%%%%%%%%%%%%%%%%%%%%%%%%%%%%%%%%%%%%%%%
\section{Neutrino Mass\label{sec:neutral_fermion_mass}}
Having described the model and introduced the scalar sector we now turn to the origin of neutrino mass. The Yukawa Lagrangian \eqref{eq:new_yukawa} mixes the SM neutrinos with the neutral fermions $\f^0_{L,R}$.
In the basis $\mathcal{V}=(\nu_L,\,\f_L^0,\,(\f_R^0)^c)^T$, we write the mass Lagrangian as
\bea
\mathcal{L}\supset -\frac{1}{2}\,\overline{\mathcal{V}^c} \,\mathcal{M} \,\mathcal{V}\ +\ \mathrm{H.c.}.
\eea
The mass matrix is comprised of two parts; namely a tree-level term and a radiative term, $\mathcal{M}=\mathcal{M}^{tree}+\mathcal{M}^{loop}$. The most important radiative correction is the contribution to the SM neutrino mass matrix, which results from the Feynman diagram in Figure~\ref{fig:loop}. One can therefore write the leading order loop-induced mass matrix as $\mathcal{M}^{loop}=\mathrm{diag}(\mathcal{M}_\nu^{loop},\,0,\,0)$. We will detail the form of $3\times3$ matrix $\mathcal{M}_\nu^{loop}$ in  Section~\ref{subsec:total_mass}; for now it suffices to note that the entries of the loop-induced mass matrix must be on the order of, or less than, the observed SM neutrino masses. In what follows we consider the seesaw and radiative masses in turn.
%%%%%%%%%%%%%%%%%%%%%%%%%%%%%%%%%%%%%%%%%%%%%%%%%%%
\subsection{Tree-Level Seesaw  Masses }

Extracting the mass terms from the Yukawa Lagrangian, one can write the mass Lagrangian as
\bea
\mathcal{L}\supset -\frac{1}{2}\,(\overline{\nu^c_L},\,\overline{(\f_L^0)^c},\,\overline{\f_R^0})
\left(\begin{array}{ccc}
\mathcal{M}_\nu^{loop}&\mathcal{M}_\s&\mathcal{M}_{\Delta}\\
\mathcal{M}_\s^T&0&\mathcal{M}_\f\\
\mathcal{M}_{\Delta}^T&\mathcal{M}_\f&0
\end{array}\right)\,
\left(\begin{array}{c}
\nu_L\\ \f_L^0\\(\f_R^0)^c
\end{array}\right),
\eea
where the Dirac mass-matrices  are
\bea
\mathcal{M}_\s\ =\ -\frac{\sqrt{3}}{2}\, \lambda_\s\,\langle S^0\rangle\,,\quad \mathcal{M}_\Delta\ =\ \frac{1}{\sqrt{3}}\, \lambda_\Delta^*\,\langle \Delta^0\rangle .\label{dirac_mixing_matrices_nu}
\eea
In general, the heavy fermion mass matrix $\mathcal{M}_\f$  is an $n\times n$ matrix for $n$ generations of exotic fermions. We consider the minimal case of $n=1$, for which $\mathcal{M}_\f=M_\f$.\footnote{Results written in terms of $\mathcal{M}_\f$ also hold for the  more general case of  $n>1$, for which one can always work in a basis where $\mathcal{M}_\f$ is diagonal, $\mathcal{M}_\f=\mathrm{diag}(M_{\f,1},\,M_{\f,2},\ldots,\,M_{\f,n})$. Note that we do not include the effects of Yukawa terms like $\overline{\f}_L\f_R\Delta$ and $\f^T_LC^{-1}\f_L S$. These are suppressed by the small VEVs, and for simplicity we assume small-enough Yukawa couplings so they can be neglected.}  The mass matrix can be partitioned into a standard seesaw form:
\bea
\mathcal{M}&=&\left(
\begin{array}{cc}
\mathcal{M}_\nu^{loop}&\mathcal{M}_D\\
\mathcal{M}_D^T&\mathcal{M}_H
\end{array}
\right),
\eea
where the Dirac mass matrix $\mathcal{M}_D$ and the heavy-fermion mass matrix $\mathcal{M}_H$ are
\bea
\mathcal{M}_D&=& (\mathcal{M}_\s\,, \mathcal{M}_{\Delta})\,,\quad
\mathcal{M}_H\ =\ \left(
\begin{array}{cc}
0&\mathcal{M}_{\f}\\
\mathcal{M}_{\f}&0
\end{array}
\right).
\eea
For $M_\f\sim$~TeV and $\langle S^0\rangle\ll\langle \Delta^0\rangle\lesssim$~GeV, the entries of the distinct mass-matrices are hierarchically separated, which we denote symbolically as:
\bea
\mathcal{M}_\nu^{loop}\ll \mathcal{M}_D\ll \mathcal{M}_\f.
\eea
With this hierarchy, a standard leading-order seesaw diagonalization can be performed. 

The  mass eigenstates are related to the interaction states via $\mathcal{V}_{\ell}=\mathcal{U}_{\ell i}\mathcal{V}_i$, where the leading-order expression for the rotation $\mathcal{U}$ is
\bea
\mathcal{U}&=&\left(\begin{array}{cc}
U_{\nu}& \mathcal{M}_D^*\,\mathcal{M}_H^{-1}\\
-\mathcal{M}_H^{-1}\,\mathcal{M}_D^T\,U_{\nu}&1
\end{array}
\right)\nonumber\\
&=&\left(\begin{array}{ccc}
U_{\nu}& \mathcal{M}_\Delta^*\,\mathcal{M}_\f^{-1}&\mathcal{M}_\s^*\,\mathcal{M}_\f^{-1}\\
-\mathcal{M}_\f^{-1}\,\mathcal{M}_\Delta^T\,U_{\nu}&1&0\\
-\mathcal{M}_\f^{-1}\,\mathcal{M}_\s^T\,U_{\nu}&0&1
\end{array}
\right).
\eea
The diagonalized  mass matrix is
\bea
\mathcal{U}^T\, \mathcal{M}\,\,\mathcal{U} &=& 
\left(\begin{array}{cc}
U_{\nu}^T(\mathcal{M}_\nu^{tree}+\mathcal{M}_\nu^{loop})U_{\nu}&0\\
0&\mathcal{M}_H
\end{array}
\right),
\eea
where $U_\nu$ is the PMNS matrix which diagonalizes the mass matrix for the light SM neutrinos:
\bea
U_{\nu}^T(\mathcal{M}_\nu^{tree}+\mathcal{M}_\nu^{loop})U_{\nu}&=&\mathrm{diag}(m_1,\,m_2,\,m_3).
\eea
The heavy neutrinos receive mass corrections of order $\mathcal{M}_\Delta\mathcal{M}_{\s}\mathcal{M}_\f^{-1}$, which split the would-be heavy Dirac fermion into a pseudo-Dirac pair. However, this splitting is tiny, being on the order of the light neutrino masses, and can be  neglected for all practical purposes. To good approximation the heavy neutrinos can be treated as a Dirac particle. 

The tree-level piece of the SM neutrino mass-matrix has a standard seesaw form:
\bea
\mathcal{M}^{tree}_\nu &=& - \mathcal{M}_D\,\mathcal{M}_{H}^{-1}\,\mathcal{M}_D^T+\mathcal{O}([\mathcal{M}_D\mathcal{M}_H^{-1}]^2)\nonumber\\
&=& -\mathcal{M}_{\Delta}\, \mathcal{M}_\f^{-1}\, \mathcal{M}^T_{\s} - \mathcal{M}_{\s}\,\mathcal{M}_\f^{-1}\,\mathcal{M}_\Delta^T+\ldots
\eea
giving
\bea
(\mathcal{M}^{tree}_\nu )_{\ell\ell'}&\simeq& \frac{1}{2}\left\{ (\lambda^*_\Delta)_{\ell } \,(\lambda_{\s})_{\ell'} + (\lambda_{\s})_{\ell } \,(\lambda^*_\Delta)_{\ell'}\right\}\frac{\langle \Delta^0\rangle\langle S^0\rangle}{M_{\f}},
\eea
where $\ell,\,\ell'\in\{e,\,\mu,\,\tau\}$ label  SM flavors. This expression has the familiar seesaw form of a Dirac-mass-squared divided by a heavy fermion mass. Using Eq.~\eqref{s_vev} one can manipulate the tree-level  mass matrix to obtain
\bea
(\mathcal{M}^{tree}_\nu )_{\ell\ell'}&\simeq& \frac{\lambda}{4}\left\{ (\lambda^*_\Delta)_{\ell } \,(\lambda_{\s})_{\ell'} + (\lambda_{\s})_{\ell } \,(\lambda^*_\Delta)_{\ell'}\right\}\frac{\langle \Delta^0\rangle^2}{M_{\s}^2}\frac{\langle H^0\rangle^2}{M_{\f}},
\eea
which will be useful in what follows. Furthermore, denoting  the beyond-SM  dimensionful parameters by a common scale $\Lambda$ and using Eq.~\eqref{delta_vev} gives
\bea
(\mathcal{M}^{tree}_\nu )_{\ell\ell'}&\simeq& \frac{\lambda}{16}\left\{ (\lambda^*_\Delta)_{\ell } \,(\lambda_{\s})_{\ell'} + (\lambda_{\s})_{\ell } \,(\lambda^*_\Delta)_{\ell'}\right\}\frac{\langle H^0\rangle^6}{\Lambda^5}.
\eea
This shows that the tree-level masses arise from a low-energy effective operator with $d=9$, giving $m_\nu\lesssim \langle H^0\rangle^6/\Lambda^5$ as expected.

%%%%%%%%%%%%%%%%%%%%%%%%%%%%%%%%%%%%%%%%%%%
%%%%%%%%%%%%%%%%%%%%%%%%%%%%%%%%%%%%%%%%%%%
\subsection{Combined Loop- and Tree-Level Masses\label{subsec:total_mass}}
In addition to the $d=9$ tree-level diagram one must calculate the $d=5$ radiative diagram  in Figure~\ref{fig:loop}. There are three  distinct diagrams with different sets of virtual fields propagating in the loops. One diagram contains the neutral fields $\{\mathcal{F}^0,\,\Delta^0,\,S^0\}$, and the other two  have the singly-charged fields $\{\f^-,\,\Delta^+,\,S^+\}$ and  $\{\mathcal{F}^+,\,\Delta^-,\,S^-\}$, respectively.\footnote{Recall that $\f^-$ is not the anti-particle of $\f^+$, and that $S^-\ne(S^+)^*$.} To good approximation one can neglect the splitting between members of a given multiplet when calculating the loop diagrams. The components of $\f$ have degenerate tree-level masses that are lifted  by radiative effects. We shall see in Section~\ref{sec:exotic_fermion_decay} that these mass-splittings are much smaller than the common tree-level mass. Similarly, the components of $S$ receive small radiative mass-splittings that can be neglected when calculating the loop-masses.\footnote{ The components of $S$ also receive tree-level splittings due to the VEVs of the various scalars. The only contributions that can be sizable  come from $\langle H^0\rangle\ne0$. However, constraints from the $\rho$ parameter require $\Delta M_\s^2\lesssim\mathcal{O}(10)$~GeV~\cite{Einhorn:1981cy}, consistent with such splittings being small.}   With this approximation, the only differences between the loop-diagrams are the numerical factors from the vertices, and the total mass-matrix for the  SM neutrinos is
\bea
\mathcal{M}_\nu&=& \mathcal{M}^{tree}_\nu\ +\  \mathcal{M}^{loop}_\nu\nonumber\\
&\simeq&\frac{\lambda}{4}\left\{ (\lambda^*_\Delta)_{\ell } \,(\lambda_{\s})_{\ell'} + (\lambda_{\s})_{\ell } \,(\lambda^*_\Delta)_{\ell'}\right\}\frac{\langle H^0\rangle^2}{M_{\f}}\times\nonumber\\
&&\left\{\frac{\langle \Delta^0\rangle^2}{M_{\s}^2}+\frac{(3\sqrt{2}-2)}{48\pi^2}\frac{M_{\f}^2}{M_{\s}^2-M_{\Delta}^2}\left[ \frac{M^2_{\s}}{M_{\mathcal{F}}^2-M_{\s}^2}\,\log \frac{M_{\mathcal{F}}^2}{M_{\s}^2}\ -\ (M_{\s}\rightarrow M_\Delta)\right]\right\}\,.\nonumber\\
& &\label{combined_nu_mass}
\eea
The tree- (loop-) mass is the first (second) term in the curly brackets. Observe that both terms have an identical flavor structure; this is a signature feature of the seesaw/radiative models, and it means the structure of the matrix that diagonalizes $\mathcal{M}_\nu$ does not depend on whether the tree- or loop-mass is dominant. Furthermore, the ratio $\mathcal{M}_\nu^{tree}/\mathcal{M}_\nu^{loop}$  is  insensitive to both  the Yukawa couplings $\lambda_{\Delta,\s}$, and the quartic coupling $\lambda$; therefore the small-$\lambda$ limit, which makes $\langle S^0\rangle$ small,   does not affect $\mathcal{M}_\nu^{tree}/\mathcal{M}_\nu^{loop}$. We present various limits of the radiative mass in Appendix~\ref{sec:loop_mass_limits}.

\begin{figure}[ttt]
\begin{center}
        \includegraphics[width = 0.65\textwidth]{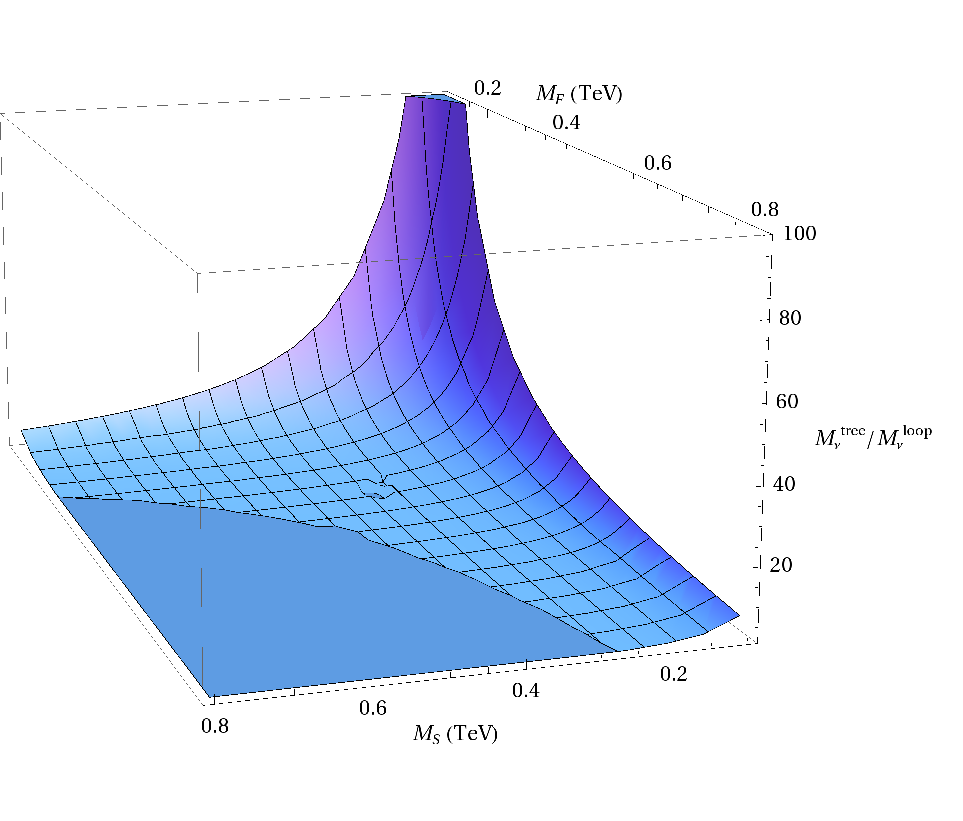}
\end{center}
\caption{Ratio of the tree-level mass to the loop-mass, $M^{tree}_\nu/M_\nu^{loop}$, as a function of the fermion ($M_\f$) and scalar ($M_\s$) masses. The meshed (plain) region has $M^{tree}_\nu/M_\nu^{loop}>1$ ($<1)$, and $M^{tree}_\nu/M_\nu^{loop}>0.1$ for the entire region.}\label{fig:mass_ratio_plot}
\end{figure}

It is interesting to determine the regions of parameter space in which the seesaw-mass dominates. Eq.~\eqref{combined_nu_mass}  shows that  larger values of $\langle\Delta^0\rangle$ tend to increase the ratio $\mathcal{M}_\nu^{tree}/\mathcal{M}_\nu^{loop}$, while larger values of $M_\Delta\gg M_\s$ tend to suppress this ratio.  We plot $\mathcal{M}_\nu^{tree}/\mathcal{M}_\nu^{loop}$ as a function of the fermion mass $M_\f$ and the scalar mass $M_\s$ in Figure~\ref{fig:mass_ratio_plot}. The fixed values $M_\Delta=7$~TeV and $\langle\Delta^0\rangle=4$~GeV are used. The plot shows the region in the $(M_\f,M_\s)$ plane of greatest interest for the LHC. The tree-level mass is dominant in much of this parameter space and the ratio satisfies $\mathcal{M}_\nu^{tree}/\mathcal{M}_\nu^{loop}>0.1$ for the entire region shown. As can be seen, keeping one of the masses small ($\lesssim300$~GeV), the other can be taken  large ($\gg\mathrm{TeV}$) while retaining $\mathcal{M}_\nu^{tree}/\mathcal{M}_\nu^{loop}>1$, while for $M_\f\sim M_\s$ the loop-mass becomes dominant for  $M_\f\sim\,M_\s\gtrsim$~TeV. Thus, for heavy $\Delta$ the tree-level region of parameter space will be most relevant for colliders like the LHC.

We are interested in the collider phenomenology of the exotic fermion $\f$ and take it as the lightest beyond-SM multiplet. We focus on the parameter space in which the triplet $\Delta$ is the heaviest, namely 
\bea
M_\f\ \lesssim\ M_\s\ \ll\ M_\Delta\,.
\eea
Within this range, the specific value of $M_\s$ is not particularly important for the collider phenomenology of $\f$. The reader should keep in mind that, in terms of the mechanism of neutrino mass, values of $M_\s$ in the lower range of this interval tend to  increase the ratio $\mathcal{M}_\nu^{tree}/\mathcal{M}_\nu^{loop}$, while larger values  have the reverse effect.

%---------------------------------------------------------
%---------------------------------------------------------

%%%%%%%%%%%%%%%%%%%%%%%%%%%%%%%%%%%%%%%%%%%%%%%%%%%%%
\section{Fixing the Yukawa Couplings\label{sec:applying_data}}

With three SM neutrinos, a generic mass matrix of the form
\bea
\mathcal{L}\supset -\frac{1}{2}\,(\overline{\nu^c_L},\,\overline{(\f_L^0)^c},\,\overline{\f_R^0})
\left(\begin{array}{ccc}
0&\mathcal{M}_\s&\mathcal{M}_{\Delta}\\
\mathcal{M}_\s^T&0&M_\f\\
\mathcal{M}_{\Delta}^T&M_\f&0
\end{array}\right)\,
\left(\begin{array}{c}
\nu_L\\ \f_L^0\\(\f_R^0)^c
\end{array}\right),
\eea
 has a vanishing determinant. The mass eigenvalues therefore contain one massless and two massive (mostly) SM neutrinos in addition to  a pseudo-Dirac heavy exotic fermion. The presence of a massless neutrino means the absolute neutrino mass scale is fixed by the observed mass-squared differences. Furthermore,  up to an overall scale factor, the couplings can be largely expressed in terms of the oscillation observables.

Writing the PMNS mixing-matrix as
\bea
U_\nu&=&
\left(\begin{array}{ccc}
c_{12}c_{13}& s_{12}c_{13}& s_{13}e^{-i\delta}\\
-s_{12}c_{23}-c_{12}s_{23}s_{13}e^{i\delta}&c_{12}c_{23}-s_{12}s_{23}s_{13}e^{i\delta}&s_{23}c_{13}\\
s_{12}s_{23}-c_{12}c_{23}s_{13}e^{i\delta}&-c_{12}s_{23}-s_{12}c_{23}s_{13}e^{i\delta}&c_{13}c_{23}
\end{array}
\right)\times U_\alpha\,,
\eea
 the matrix $U_\alpha$ contains the Majorana phase, and can be taken as $U_\alpha=\mathrm{diag}(e^{-i\alpha},\,e^{i\alpha},\,1)$ for our case with a massless neutrino~\cite{Gavela:2009cd}. The best-fit neutrino oscillation parameters are listed in Table~\ref{neutrino_mass_data_table}~\cite{Schwetz:2012yy}, and we use these for our numerics throughout. The CP phase $\delta$ and the Majorana phase $\alpha$ are not experimentally known and can assume any value  at the 2$\sigma$ level; we therefore treat these as free parameters.

We denote the ratio of mass-squared differences by
\bea
r&=&\frac{|\Delta m_{12}^2|}{|\Delta m_{13}^2|}\ \ll\ 1,
\eea
and write the Yukawa couplings as
\bea
\lambda_\s\ \equiv\ r_\s\,\hat{\lambda}_\s\,\quad \mathrm{and}\quad\lambda_\Delta\ \equiv\ r_\Delta\,\hat{\lambda}_\Delta.
\eea
Here $r_{\s,\Delta}$ are the magnitudes of the flavor-space vectors $\lambda_{\s,\Delta}$, so that  $\hat{\lambda}_\s$ and $\hat{\lambda}_\Delta$ are complex vectors of unit norm:
\bea
\sum_\ell(\hat{\lambda}_{\s}^*)_{\ell}\,(\hat{\lambda}_{\s})_{\ell}=1\quad\mathrm{and}\quad\sum_\ell(\hat{\lambda}_{\Delta}^*)_{\ell}\,(\hat{\lambda}_{\Delta})_{\ell}=1.
\eea
In the following we obtain the form of these unit vectors for a normal hierarchy and an inverted hierarchy. This information influences the  collider signals of the model.

%%%%%%%%%%%%%%%%%%%%%%%%%%%%%%%%%%%%%%%%%%%%%%%%%%%%%%%%
\begin{table}
\centering
\begin{tabular}{|ccc|}\hline
&\multicolumn{2}{|c|}{}\\
Parameter& \multicolumn{2}{|c|}{Best fit ($\pm1\sigma$)} \\
&\multicolumn{2}{|c|}{}\\\cline{2-3}
&\multicolumn{1}{|c|}{Normal Hierarchy} &\multicolumn{1}{|c|}{Inverted Hierarchy}\\
\hline
%-----------------------------------------------%
&&\\
$\Delta m_{12}^2\;(10^{-5}~\mathrm{eV}^2)$ &
$7.59^{+0.20}_{-0.18}$ & $7.59^{+0.20}_{-0.18}$ \\
& & \\
$\Delta m_{31}^2\;(10^{-3}~\mathrm{eV}^2)$ &
$2.50^{+0.09}_{-0.16}$ & $-2.40^{+0.08}_{-0.09}$ \\
& & \\
$\sin^2\theta_{12}$ &
$0.312^{+0.017}_{-0.015}$ &$0.312^{+0.017}_{-0.015}$ \\
& & \\
$\sin^2\theta_{23}$ &
$0.52^{+0.06}_{-0.07}$ &$0.52^{+0.06}_{-0.06}$ \\
& & \\
$\sin^2\theta_{13}$ &
$0.013^{+0.007}_{-0.005}$ &$0.016^{+0.008}_{-0.006}$ \\
& & \\
\hline
%------------------------------------------------%
\end{tabular}
\caption{\label{neutrino_mass_data_table} Neutrino oscillation parameters~\cite{Schwetz:2012yy}.}
\end{table}
%%%%%%%%%%%%%%%%%%%%%%%%%%%%%%%%%%%%%%%%%%%%%%%%%%%%%
%%%%%%%%%%%%%%%%%%%%%%%%%%%%%%%%%%%%%%%%%%%%%%%%%%%%%
\subsection{Normal Hierarchy}
Consider a region of parameter space in which the the seesaw mass is dominant and the radiative mass can be neglected. Defining the quantity $R_N$ as
\bea
R_N\ =\ \frac{\sqrt{1+r}-\sqrt{r}}{\sqrt{1+r}+\sqrt{r}},
\eea
the results of Ref.~\cite{Gavela:2009cd} allow one to write the mass eigenvalues as
\bea
m_1&=&0\,,\nonumber\\
|m_2|&=&\frac{\,r_\s\,r_\Delta}{2}\,\frac{\langle\Delta^0\rangle\langle S^0\rangle}{M_\f}\times (1-R_N)\,,\nonumber\\
  |m_3|&=&\frac{\,r_\s\,r_\Delta}{2}\,\frac{\langle\Delta^0\rangle\langle S^0\rangle}{M_\f}\times (1+R_N)\,.\label{eq:NH_masses}
\eea
Noting that
\bea
|m_2|&=& \sqrt{|\Delta m_{21}^2|}\ \approx\ \sqrt{7.59\times 10^{-5}}~\mathrm{eV},\nonumber\\
|m_3|&=& \sqrt{|\Delta m_{31}^2|}\ \approx\ \sqrt{2.5\times 10^{-3}}~\mathrm{eV},
\eea
 fixes the product of parameters appearing  in Eq.~\eqref{eq:NH_masses} as
\bea
\frac{|m_2|}{(1-R_N)}&=&\frac{|m_3|}{(1+R_N)}\ \approx\ 0.029~\mathrm{eV}.\label{mass_ex_values_NH}
\eea
Due to the related flavor dependence of the seesaw and radiative masses,  one can extend these results to the more general case where the radiative mass is  important/dominant. Rewriting Eq.~\eqref{eq:NH_masses} to include the loop mass in Eq.~\eqref {combined_nu_mass} gives
\bea
\frac{|m_2|}{(1-R_N)}&=&\frac{|m_3|}{(1+R_N)}\ \simeq\ r_\s\,r_\Delta\,\frac{|\lambda|}{4}\,\frac{\langle H^0\rangle^2}{M_\f}\left\{\frac{\langle\Delta^0\rangle^2}{M_\s^2}\ +\ \mathrm{loop\ piece\ }\right\}.\label{eq:mass_scale_wLoop}
\eea
Combining Eqs.~\eqref{eq:mass_scale_wLoop}  and~\eqref{mass_ex_values_NH} allows one to fix a combination of the parameters in terms of the overall scale of the neutrino masses in the general case.

The Yukawa unit vectors can be fixed in terms of the oscillation observables:
\bea
\hat{\lambda}_{\s,\ell}&=& \frac{1}{\sqrt{2}}\,\left(\sqrt{1+R_N} \,(U_\nu^*)_{\ell3} +\sqrt{1+R_N}\,(U_\nu^*)_{\ell2} \right)\,,\nonumber\\
\hat{\lambda}_{\Delta,\ell}&=& \frac{1}{\sqrt{2}}\,\left(\sqrt{1+R_N} \,(U_\nu)_{\ell3} -\sqrt{1+R_N}\,(U_\nu)_{\ell2} \right)\,,
\eea
up to the dependence on the unknown phases $\delta$ and $\alpha$. For example, with $\delta=\alpha=0$, the PMNS best-fit values give
\bea
\hat{\lambda}_{\s}&\approx&  (0.32,\,0.86,\,0.39)^T\,,\nonumber\\
\hat{\lambda}_{\Delta}&\approx&(-0.10,\, 0.46,\,0.88)^T\,.
\eea

 If the tree-level mass is dominant, or on the order of the loop-mass, one has
\bea
\frac{|m_{2,3}|}{(1\mp R_N)}&\approx& \left(\frac{\sqrt{3}}{2}r_s\,\langle S^0\rangle\right) \left(\frac{1}{\sqrt{3}}\,r_\Delta\,\langle \Delta^0\rangle\right)\,\frac{1}{M_\f}\ \equiv\  |\mathcal{K}_\s|\,|\mathcal{K}_\Delta| \,M_\f,\label{eq:relating_Ks}
\eea
where we introduce  the dimensionless vectors 
\bea
\mathcal{K}_{\s,\Delta}\ =\  \mathcal{M}_{\s,\Delta}\,\mathcal{M}^{-1}_\f,
\eea 
and  write their magnitude as $|\mathcal{K}_{\s,\Delta}|$. We shall see later that the quantities $\mathcal{K}_{\s,\Delta}$  influence the decay properties of the exotic fermions. Eqs.~\eqref{mass_ex_values_NH} and ~\eqref{eq:relating_Ks} give the relation
\bea
|\mathcal{K}_{\s}|&\approx& 5.8\times 10^{-7} \times\sqrt{\frac{\mathrm{GeV}}{M_\f}} \times\left\{\frac{10^{-6} \sqrt{2500/(M_\f/\mathrm{GeV})}}{|\mathcal{K}_{\Delta}|}\right\},
\eea
for the case of a normal hierarchy. This fixes the relative size of $|\mathcal{K}_{\s}|$ and $|\mathcal{K}_{\Delta}|$ in terms of the overall scale of the SM neutrino masses. Provided the radiative mass is not significantly larger than the tree-level mass, this relationship also provides a good approximation in the presence of the loop-mass.

%%%%%%%%%%%%%%%%%%%%%%%%%%%%%%%%%%%%%%%%%%%%%%%%%%%%%
\subsection{Inverted Hierarchy}
The same procedure can be followed for the inverted hierarchy; defining 
\bea
R_I\ =\ \frac{\sqrt{1+r}-1}{\sqrt{1+r}+1},
\eea
the light-neutrino mass eigenvalues are
\bea
|m_1|&=&\frac{1}{2}\,r_\s\,r_\Delta\,\frac{\langle\Delta^0\rangle\langle S^0\rangle}{M_\f}\times (1-R_I)\,,\nonumber\\
  |m_2|&=&\frac{1}{2}\,r_\s\,r_\Delta\,\frac{\langle\Delta^0\rangle\langle S^0\rangle}{M_\f}\times (1+R_I)\,,\nonumber\\
m_3&=&0,
\eea
when the tree-level mass dominates. The nonzero mass-eigenvalues are fixed via the observed mass-squared differences,
 giving
\bea
\frac{|m_1|}{(1-R_I)}&=&\frac{|m_2|}{(1+R_N)}\ \approx\ 0.049~\mathrm{eV}.\label{eq:IH_mass_values}
\eea

Including the loop mass gives 
\bea
|m_1|&=&r_\s\,r_\Delta\,\frac{|\lambda|}{4}\,\frac{\langle H^0\rangle^2}{M_\f}\left\{\frac{\langle\Delta^0\rangle^2}{M_\s^2}\ +\ \mathrm{loop\ piece\ }\right\}\times (1-R_I)\,,\nonumber\\
  |m_2|&=&r_\s\,r_\Delta\,\frac{|\lambda|}{4}\,\frac{\langle H^0\rangle^2}{M_\f}\left\{\frac{\langle\Delta^0\rangle^2}{M_\s^2}\ +\ \mathrm{loop\ piece\ }\right\}\times (1+R_I)\,,
\eea
allowing one to fix the overall scale via \eqref{eq:IH_mass_values}.  The Yukawa unit-vectors are fixed to be
\bea
\hat{\lambda}_{\s,\ell}&=&\frac{1}{\sqrt{2}}\,\left(\sqrt{1+R_I} \,(U_\nu^*)_{\ell2} +\sqrt{1+R_I}\,(U_\nu^*)_{\ell1} \right)\,,\nonumber\\
\hat{\lambda}_{\Delta,\ell}& =& \frac{1}{\sqrt{2}}\,\left(\sqrt{1+R_I} \,(U_\nu)_{\ell2} -\sqrt{1+R_I}\,(U_\nu)_{\ell1} \right)\,.
\eea
When the tree-level mass is dominant, or on the order of the loop-mass, one has
\bea
\frac{|m_{1,2}|}{(1\mp R_I)}&\approx& \left(\frac{\sqrt{3}}{2}r_s\,\langle S^0\rangle\right) \left(\frac{1}{\sqrt{3}}\,r_\Delta\,\langle \Delta^0\rangle\right)\,\frac{1}{M_\f}\ =\ |\mathcal{K}_\s|\,|\mathcal{K}_\Delta| \,M_\f.
\eea
 giving the relation
\bea
|\mathcal{K}_{\s}|&\approx& 3.4\times 10^{-7} \times\sqrt{\frac{\mathrm{GeV}}{M_\f}} \times\left\{\frac{10^{-6} \sqrt{2500/(M_\f/\mathrm{GeV})}}{|\mathcal{K}_{\Delta}|}\right\},
\eea
for the inverted hierarchy.
%%%%%%%%%%%%%%%%%%%%%%%%%%%%%%%%%%%%%%%%%%%
%%%%%%%%%%%%%%%%%%%%%%%%%%%%%%%%%%%%%%%%%%%
%%%%%%%%%%%%%%%%%%%%%%%%%%%%%%%%%%%%%%%%%%%
\section{Collider Production of Exotic Fermions \label{sec:production_of_fermions}}
We now consider the production of exotic fermions at the LHC. The exotics are most readily produced in pairs via electroweak interactions.\footnote{This differs from the composite-fermion model of Ref.~\cite{Biondini:2012ny}, which has a sizable $W\ell^+ \f^{--}$ vertex and gives single-fermion production, $q\bar{q}'\rightarrow \overline{\f^{--}}\ell$. In our context pair-production is favored due to the Yukawa suppression of, e.g., the $W\ell^+ \f^{--}$ coupling when $M_\f\sim$~TeV.} The interactions of $\mathcal{F}$ with electroweak gauge bosons arise from the kinetic Lagrangian
\bea
\mathcal{L}\supset i\overline{\mathcal{F}}D_\mu\gamma^\mu \mathcal{F},
\eea
where the covariant derivative has the standard form
\bea
D_\mu\mathcal{F}=\left[\partial_\mu -i gT_a W^a_\mu-ig'\frac{Y}{2}B_\mu\right]\mathcal{F}.
\eea
Writing the fermion as 
\bea
\mathcal{F}=(\f^+,\,\f^0,\,\f^-,\,\f^{--})^T\,,
\eea
a suitable set of generators $T_a$ for the quadruplet (isospin-3/2) representation of $SU(2)$ is
\bea
T_1&=&\frac{1}{2}\left(
\begin{array}{cccc}
0&\sqrt{3}&0&0\\
\sqrt{3}&0&2&0\\
0&2&0&\sqrt{3}\\
0&0&\sqrt{3}&0
\end{array}
\right),\quad
T_2\ =\ \frac{1}{2}\left(
\begin{array}{cccc}
0&-i\sqrt{3}&0&0\\
i\sqrt{3}&0&-2i&0\\
0&2i&0&-\sqrt{3}i\\
0&0&\sqrt{3}i&0
\end{array}
\right),\nonumber\\
\nonumber\\
&&\qquad\qquad \qquad\  T_3=\mathrm{diag}(3/2,\,1/2,\,-1/2,\,-3/2)\,.
\eea
The interaction Lagrangian contains the terms
\bea
\mathcal{L}&\supset& g\left\{\sqrt{\frac{3}{2}}\,\overline{\f^{+}}\gamma^\mu\f^0+\sqrt{2}\,\overline{\f^{0}}\gamma^\mu \f^-+\sqrt{\frac{3}{2}}\,\overline{\f^-}\gamma^\mu \f^{--}\right\}W^+_\mu\nonumber\\
& & + g\left\{\sqrt{\frac{3}{2}}\,\overline{\f^{0}}\gamma^\mu\f^{+}+\sqrt{2}\,\overline{\f^-}\gamma^\mu\f^0+\sqrt{\frac{3}{2}}\,\overline{\f^{--}}\gamma^\mu \f^-\right\}W^-_\mu\label{eq:int_lagrangian}\\
& & + e\left\{\overline{\f^{+}}\gamma^\mu\f^+-\,\overline{\f^-}\gamma^\mu \f^--2\,\overline{\f^{--}}\gamma^\mu\f^{--}\right\}A_\mu\nonumber\\
& &+ \frac{g}{c_{\lw}}\left\{g_{\lz\lf^{+}}\overline{\f^{+}}\gamma^\mu\f^++g_{\lz\lf^{0}}\overline{\f^0}\gamma^\mu \f^0+g_{\lz\lf^{-}}\,\overline{\f^-}\gamma^\mu \f^-+g_{\lz\lf^{--}}\,\overline{\f^{--}}\gamma^\mu\f^{--}\right\}Z_\mu\,,\nonumber
\eea
where the couplings to the $Z$-boson have the standard form, $g_{\lz\lf}=(I_3^{\lf}-s_{\lw}^2Q_{\lf})$, giving
\bea
g_{\lz\lf^{+}}=\frac{1}{2}\left(3 - 2s_{\lw}^2\right)\;,\quad  g_{\lz\lf^{0}}=\frac{1}{2}\;,\quad  g_{\lz\lf^{-}}=\frac{1}{2}\left(-1 + 2s_{\lw}^2\right)\,,\quad g_{\lz\lf^{--}}=\frac{1}{2}\left(-3 +4s_{\lw}^2\right) .\nonumber
\eea
We define $g_{\lw\lf}=\sqrt{3/2}$ or $g_{\lw\lf}=\sqrt{2}$, depending on which pairs of fermions the $W$ is coupling to (the correct choice can be read off the Lagrangian above).

The partonic-level cross section for pair production of the exotic fermions may be written as
\bea
\hat{\sigma}(q\bar{q}\rightarrow \overline{\f^{\ }}\f;\hat{s})&=& \frac{\beta(3-\beta^2)}{48\pi^2}\, \hat{s}\, (|V_L|^2+|V_R|^2),
\eea
where $\hat{s}=(p_q+p_{\bar{q}})^2$ is the usual Mandelstam variable, evaluated for the incoming partons $q$ and $\bar{q}$. The exotic-fermion velocity is denoted by $\beta=\sqrt{1-4M_{\f}^2/\hat{s}}$, and the couplings $V_{L,R}$ have the form
\bea
V_{L,R}^{\gamma+Z}&=&\frac{Q_{\lf}\,Q_q\,e^2}{\hat{s}}+ \frac{g_{\lz\lf}\,g_{L,R}^q\,g^2}{c_{\lw}^2\,(\hat{s}-M_{\lz}^2)}\;,\\
V_{L}^{W^-}&=&\left[V_L^{W^+}\right]^*\ =\ \frac{g_{\lw\lf}\,g^2\,V_{ud}}{\sqrt{2}\,(\hat{s}-M_{\lw}^2)}.
\eea
The right-chiral coupling for the $W$ boson vanishes, $V_R^{W^{\pm}}=0$, and the quark couplings are $g^q_{L,R}=T_3^q-s_{\lw}^2Q_q$.

%---------------------------------------------------------
\begin{figure}[ttt]
\centering
\begin{tabular}{cc}
\epsfig{file=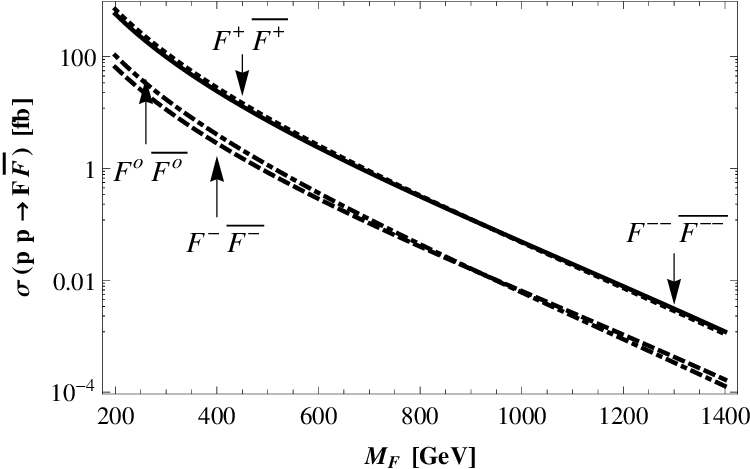,width=0.5\linewidth,clip=}&
\epsfig{file=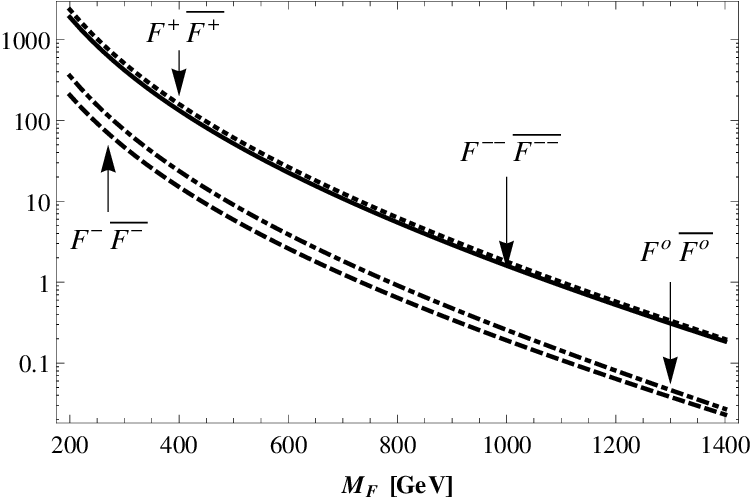,width=0.47\linewidth,clip=}
\end{tabular}
\caption{Neutral-current production cross section for exotic fermion pairs at the LHC. The left (right) plot is for $\sqrt{s}=7~(14)$~TeV. }\label{fig:pp_Z_production}
\end{figure}
%----------------------------------------------------

The partonic-level cross section must be convoluted with an appropriate parton distribution function (PDF) to determine the  cross section for a hadron collider. This gives
\bea
\sigma(pp\rightarrow \overline{\f}\f)&=&\frac{1}{N_c}\sum_{q=u,d,\ldots}\sum_{\bar{q}=\bar{u},\bar{d},\ldots}\int_0^1dx_1\int_0^1dx_2\,\hat{\sigma}(q\bar{q}\rightarrow \overline{\f^{\ }}\f;\hat{s}=x_1x_2s)\nonumber\\
& &\times\left\{f_{q/p}(x_1,\mu^2)f_{\bar{q}/p}(x_2,\mu^2)+f_{q/p}(x_2,\mu^2)f_{\bar{q}/p}(x_1,\mu^2)\right\}\,,
\eea
where $f_{q/p}(x,\mu^2)$ are the PDFs, for which we employ the MSTW08 set~\cite{Martin:2009iq} with the factorization scale set at $\mu^2=M_\f^2$. Here $\sqrt{s}$ is the $pp$ beam energy and the  color pre-factor accounts for the average over initial-state colors.  The cross section  is  multiplied by the standard initial-state $K$-factor~\cite{Hamberg:1990np},
\bea
K_i(q^2)&\approx&1+\frac{\alpha_s(q^2)}{2\pi}\,\frac{4}{3}\,(1+\frac{4\pi^2}{3}),
\eea
to incorporate QCD corrections.

\begin{figure}[ttt]
\centering
\begin{tabular}{cc}
\epsfig{file=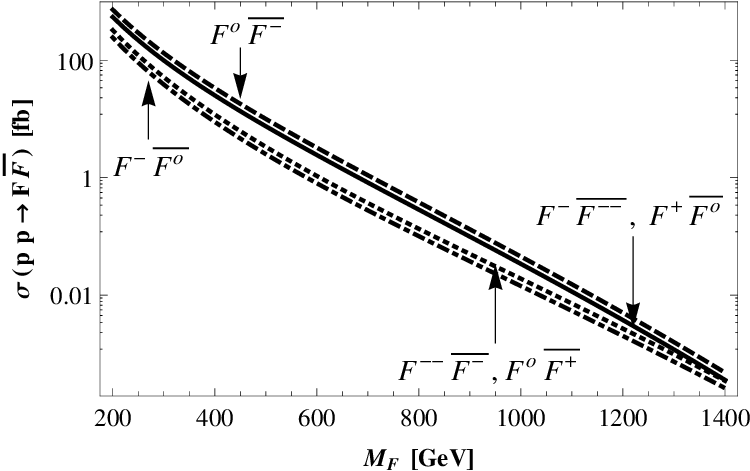,width=0.49\linewidth,clip=}&
\epsfig{file=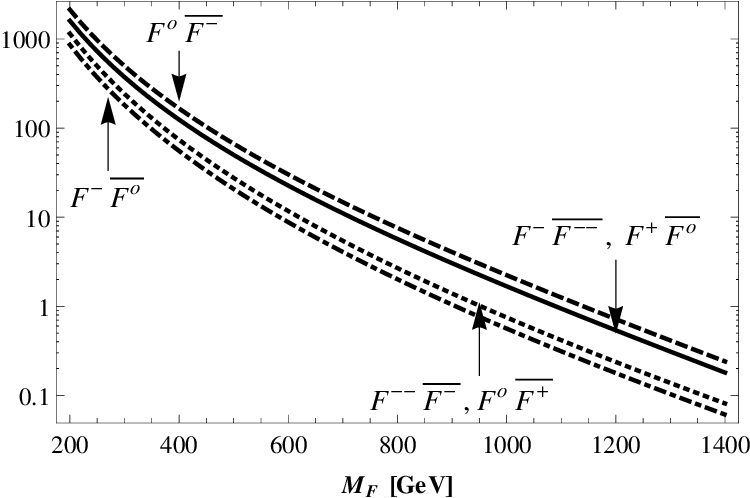,width=0.47\linewidth,clip=}
\end{tabular}
\caption{Charged-current production cross section for exotic fermion pairs at the LHC. The left (right) plot is for $\sqrt{s}=7~(14)$~TeV. }\label{fig:pp_W_production}
\end{figure}

Let us consider which final-states are possible in $q\bar{q}\rightarrow \overline{\f^{\ }}\f$ processes. The neutral-current process $q\bar{q}\rightarrow \overline{\f^{0}}\f^0$ is mediated by the $Z$ boson, while the final states $\overline{\f^{+}}\f^+$, $ \overline{\f^{-}}\f^-$, or $ \overline{\f^{--}}\f^{--}$, can all be realized via  an intermediate $Z$ or a photon. We plot the  LHC production cross sections for these processes in Figure~\ref{fig:pp_Z_production} for $\sqrt{s}=7$~TeV and $14$~TeV.  Production via  an intermediate $W$ boson, on the other hand,  gives the final states $\overline{\f^0}\f^{+}$, $\overline{\f^{-}}\f^{0}$, or $\overline{\f^{--}}\f^{-}$ (for $W^+$), and $\overline{\f^{+}}\f^0$, $\overline{\f^{0}}\f^{-}$, or $\overline{\f^{-}}\f^{--}$ (for $W^-$). The analogous LHC production cross sections are plotted in Figure~\ref{fig:pp_W_production}.

Note that, once the beam energy is specified, the production cross section for $\f\overline{\f}$ pairs depends on the single free parameter $M_\f$.  As seen in the plots, the cross sections have typical weak-interaction  values; for an LHC operating energy of $\sqrt{s}=7$~TeV the total production cross section for  $\f\overline{\f}$ pairs is $\mathcal{O}(10^2)$~fb for $M_\f\sim350$~GeV. At higher operating energies of $\sqrt{s}=14$~TeV this increases to $\mathcal{O}(10^3)$~fb for the same value of $M_\f$. Thus, integrated luminosities of $\sim5$~fb$^{-1}$ should yield $\sim500$ (5000) production events in this mass range for $\sqrt{s}=7~(14)$~TeV. These values are typical for the $d>5$ models in Table~\ref{table:seesaw_models}~\cite{Picek:2009is,Kumericki:2012bh}; fermion production cross-sections are determined for a given value of $M_\f$, with values typical of those for weak interactions. 

%---------------------------------------------------------
%----------------------------------------------------

The production cross-sections for $\f$-pairs are insensitive to the details of the underlying neutrino-mass mechanism. For a given fixed $M_\f$ they are determined once the quantum numbers for $\f$ are specified, independent of any connection to neutrino mass. To connect the exotic fermions to the mechanism of neutrino mass one must study their decay properties. We shall see that these are sensitive to the details of the neutrino sector.

%---------------------------------------------------------
%%%%%%%%%%%%%%%%%%%%%%%%%%%%%%%%%%%%%%%%%%%
\section{Mass Eigenstate Interactions\label{sec:mass_eigenstate_ints}}
 The relationship between the exotics and neutrino mass is encoded in the Yukawa Lagrangian, which induces mass-mixing between $\f$ and SM leptons,  thereby influencing the decay channels and  branching fractions.  The mixing in the neutral-fermion sector was detailed in Section~\ref{sec:neutral_fermion_mass}; here we account for the mass mixing amongst charged fermions, then determine the mass-eigenstate interaction Lagrangian. 

The charged fermions have the following mass Lagrangian:
\bea
\mathcal{L}&\supset& -\, (\overline{\ell_R},\, \overline{\f^-_R},\,\overline{(\f^+_L)^c})
\left(\begin{array}{ccc}
\mathcal{M}_\ell&0&0\\
m_\Delta&\mathcal{M}_{\f}&0\\
m_{\s}&0&\mathcal{M}_{\f}^T
\end{array}
\right)
\left(
\begin{array}{c}
\ell_L\\
\f^-_L\\
(\f^+_R)^c
\end{array}
\right)-\mathcal{M}_{\f} \overline{\f_R^{--}}\,\f_L^{--}+\mathrm{H.c.},\nonumber
\eea
where $\mathcal{M}_\ell$ ($\mathcal{M}_{\f}$) is a diagonal mass matrix for the SM leptons (singly-charged exotics).\footnote{For a single generation of exotics one has $\mathcal{M}_\f=M_\f$.} The mixing matrices  have the form
\bea
m_\Delta=\frac{\lambda_\Delta^\dagger}{\sqrt{3}}\,\langle \Delta^0\rangle\,\quad\mathrm{and}\quad m_{\s}=-\frac{\lambda_{\s}^T}{2}\,\langle S^0\rangle\,.
\eea
The mass matrix is diagonalized via a bi-unitary transformation
\bea
U_R^\dagger 
\left(\begin{array}{ccc}
\mathcal{M}_\ell&0&0\\
m_\Delta&\mathcal{M}_{\f}&0\\
m_{\s}&0&\mathcal{M}_{\f}^T
\end{array}
\right)
U_L &\approx &\mathrm{diag}(\,\mathcal{M}_\ell,\, \mathcal{M}_{\f},\,\mathcal{M}_{\f}^T),
\eea
where we anticipate the fact that the corrections to the charged-lepton mass matrix are on the order of the SM neutrino masses and  can be neglected for all practical purposes. To leading order the rotation matrices take the form
\bea
U_L&=&\left(\begin{array}{ccc}
1&m_\Delta^\dagger \,\mathcal{M}_\f^{-1}&m_\s^\dagger \,\mathcal{M}_\f^{-1}\\
-\mathcal{M}_\f^{-1}\,m_\Delta&1&0\\
-\mathcal{M}_\f^{-1}\,m_{\s}&0&1
\end{array}
\right),\nonumber\\
U_R&=&\left(\begin{array}{ccc}
1&\mathcal{M}_\ell\,m_\Delta^\dagger \,\mathcal{M}_\f^{-2}&\mathcal{M}_\ell\,m_\s^\dagger \,\mathcal{M}_\f^{-2}\\
-\mathcal{M}_\f^{-2}\,m_\Delta\,\mathcal{M}_\ell&1&0\\
-\mathcal{M}_\f^{-2}\,m_{\s}\,\mathcal{M}_\ell&0&1
\end{array}
\right).
\eea

The mass-mixing couples the exotics and the SM leptons via the charged and neutral currents. The singly-charged fermion interaction-eigenstates have the following couplings to the $Z$ boson:
\bea
\mathcal{L}_{Z,\pm}=\frac{g}{c_{\lw}}\left\{ g_{\lz\lf^+}\overline{\f^+}\gamma^\mu \f^++g_{\lz\lf^-}\overline{\f^-}\gamma^\mu \f^- +g^\ell_{L}\overline{\ell_{L}}\gamma^\mu\ell_{L} + g^\ell_{R}\overline{\ell_{R}}\gamma^\mu\ell_{R} \right\}Z_\mu.
\eea
Rotating to the mass-basis, the interactions between mass eigenstates and the $Z$ boson are
\bea
\mathcal{L}_{Z,\pm}&=&\frac{g}{c_{\lw}}\left\{ \frac{}{}g_{\lz\lf^+}\overline{\f^+}\gamma^\mu \f^++g_{\lz\lf^-}\overline{\f^-}\gamma^\mu \f^- +g^\ell_{L}\overline{\ell_{L}}\gamma^\mu\ell_{L} + g^\ell_{R}\overline{\ell_{R}}\gamma^\mu\ell_{R} \right.\nonumber\\
& & \left.+(g^\ell_L-g_{\lz\lf^-}) \,(\overline{\ell_L}m_\Delta^\dagger \mathcal{M}_\f^{-1} \gamma^\mu \f^-_L+ \overline{\f^-_L}\mathcal{M}_\f^{-1}m_\Delta \gamma^\mu\ell_L)\right.\nonumber\\
& &\left. - (g^\ell_L+g_{\lz\lf^+})\, (\overline{\ell_L^c}\,m_\s^T \,\mathcal{M}_\f^{-1}\, \gamma^\mu \,\f^+_R+ \overline{\f^+_R}\,\mathcal{M}_\f^{-1}\,m_\s^* \,\gamma^\mu\,\ell_L^c)\right\}Z_\mu.
\eea
Using the specific values of the coupling constants gives
\bea
\mathcal{L}_{Z,\pm}&=&\frac{g}{c_{\lw}}\left\{ \frac{}{}g_{\lz\lf^+}\overline{\f^+}\gamma^\mu \f^++g_{\lz\lf^-}\overline{\f^-}\gamma^\mu \f^- +g^\ell_{L}\overline{\ell_{L}}\gamma^\mu\ell_{L} + g^\ell_{R}\overline{\ell_{R}}\gamma^\mu\ell_{R} \right.\nonumber\\
& & \left.-  \ (\overline{\ell_L^c}\,m_\s^T \,\mathcal{M}_\f^{-1}\, \gamma^\mu \,\f^+_R+ \overline{\f^+_R}\,\mathcal{M}_\f^{-1}\,m_\s^* \,\gamma^\mu\,\ell_L^c)\right\}Z_\mu.
\eea
Observe that the leading-order coupling between $\ell_L$ and $\f^-_L$ has canceled out, due to the relation $g_L^\ell=g_{\lz\lf^-}$. 

Similarly, the neutral-fermion interaction-eigenstates couple to the $Z$ boson as follows
\bea
\mathcal{L}_{Z,0}=\frac{g}{c_{\lw}}\left\{ g_{\lz\lf^0}\overline{\f^0}\gamma^\mu \f^0+g_{\nu}\overline{\nu_L}\gamma^\mu \nu_L \right\}Z_\mu.
\eea
Rotating to the mass basis gives
\bea
\mathcal{L}_{Z,0}&=&\frac{g}{c_{\lw}}\left\{ \frac{}{}g_{\lz\lf^0}\overline{\f^0}\gamma^\mu \f^0+g_{\nu}\overline{\nu_L}\gamma^\mu \nu_L \right.\nonumber\\
& &\left. +(g_\nu-g_{\lz\lf^0}) \,(\overline{\nu_L}\,U^\dagger_{\nu}\,\mathcal{M}_\Delta^*\, \mathcal{M}_\f^{-1} \,\gamma^\mu \,\f^0_L + \overline{\f^0_L}\,\mathcal{M}_\f^{-1}\,\mathcal{M}_\Delta ^T\,\gamma^\mu\,U_{\nu}\,\nu_L)\right.\nonumber\\
& &\left. - (g_\nu+g_{\lz\lf^0})\, (\overline{\nu_L^c}\,U^T_{\nu}\,\mathcal{M}_\s\, \,\mathcal{M}_\f^{-1}\, \gamma^\mu \,\f^0_R+ \overline{\f^0_R}\,\mathcal{M}_\f^{-1}\,\mathcal{M}_\s^\dagger \,\gamma^\mu\,U^*_{\nu}\,\nu_L^c)\right\}Z_\mu,\nonumber
\eea
and employing the specific values of the couplings reduces this to
\bea
\mathcal{L}_{Z,0}&=&\frac{g}{c_{\lw}}\left\{ \frac{}{}g_{\lz\lf^0}\overline{\f^0}\gamma^\mu \f^0+g_{\nu}\overline{\nu_L}\gamma^\mu \nu_L \right.\nonumber\\
& &\left.- \, (\overline{\nu_L^c}\,U^T_{\nu}\,\mathcal{M}_\s\, \,\mathcal{M}_\f^{-1}\, \gamma^\mu \,\f^0_R+ \overline{\f^0_R}\,\mathcal{M}_\f^{-1}\,\mathcal{M}_\s^\dagger \,\gamma^\mu\,U^*_{\nu}\,\nu_L^c)\right\}Z_\mu,\nonumber
\eea
where once again a cancellation has occurred.

The interaction Lagrangian for the $W$ boson and the leptons is
\bea
\mathcal{L}_{W}&=& g\left\{\sqrt{\frac{3}{2}}\,\overline{\f^{+}}\gamma^\mu\f^0+\sqrt{2}\,\overline{\f^{0}}\gamma^\mu \f^-+\sqrt{\frac{3}{2}}\,\overline{\f^-}\gamma^\mu \f^{--}+\frac{1}{\sqrt{2}}\overline{\nu_L}\gamma^\mu \ell_L\right\}W^+_\mu+\mathrm{H.c.}\nonumber
\eea
Rotating to the mass basis gives a somewhat complicated expression,
\bea
\mathcal{L}_{W}&=& g\left\{\sqrt{\frac{3}{2}}\,\overline{\f^{+}}\gamma^\mu\f^0+\sqrt{2}\,\overline{\f^{0}}\gamma^\mu \f^-+\sqrt{\frac{3}{2}}\,\overline{\f^-}\gamma^\mu \f^{--}+\frac{1}{\sqrt{2}}\overline{\nu_L}\,U^\dagger_\nu\gamma^\mu \ell_L\right\}W^+_\mu\nonumber\\
& &+g\left\{-\sqrt{\frac{3}{2}}\,\overline{\ell_L}\,m_\Delta^\dagger\,\mathcal{M}_\f^{-1}\,\gamma^\mu\,\f_L^{--}+\frac{1}{\sqrt{2}}\overline{\f_L^0}\,\mathcal{M}_\f^{-1}\left[\mathcal{M}^T_\Delta -2 m_\Delta\right]\gamma^\mu\,\ell_L\right.\nonumber\\
& & +\frac{1}{\sqrt{2}}\,\overline{\nu_L}\,U_\nu^\dagger\left[m_\Delta^\dagger -2\mathcal{M}_\Delta^*\right]\mathcal{M}_\f^{-1}\,\gamma^\mu\,\f^-_L-\frac{1}{\sqrt{2}}\,\overline{\ell^c_L}\left[\sqrt{3}\,m_\s^T+\mathcal{M}_\s\right]\mathcal{M}_\f^{-1}\,\gamma^\mu\,\f_R^0\nonumber\\
& &-\frac{1}{\sqrt{2}}\,\overline{\f^+_R}\,\mathcal{M}_\f^{-1}\left[\sqrt{3}\mathcal{M}_\s^\dagger+m_\s^*\right]\gamma^\mu\,U_\nu^*\,\nu_L^c-\sqrt{2}\,\overline{\nu_L^c}\,U_\nu^T\,\mathcal{M}_\s\,\mathcal{M}^{-1}_\f\,\gamma^\mu\,\f^-_R\nonumber\\
& &\left.-\sqrt{\frac{3}{2}}\,\overline{\f^+_L}\,\gamma^\mu\,\mathcal{M}_\f^{-1}\, \mathcal{M}_\Delta^T\,U_\nu\,\nu_L\right\}W_\mu^++\mathrm{H.c.}
\eea

These expressions for the charged- and neutral-current interactions are simplified by noting the relationship between the mass-mixing matrices in the charged and neutral sectors:
\bea
m^T_{\Delta}\ = \ \mathcal{M}_\Delta \qquad\mathrm{and}\qquad m_\s\ =\ \frac{1}{\sqrt{3}}\,\mathcal{M}_\s^T,
\eea
and recalling the definition of the matrix-valued quantities:
\bea
\mathcal{K}_{\s,\Delta}\ =\  \mathcal{M}_{\s,\Delta}\,\mathcal{M}^{-1}_\f,
\eea 
where we suppress the flavor index: $\mathcal{K}_{\s,\Delta}\equiv \mathcal{K}_{\s,\Delta}^\ell$. The $Z$ boson interaction-Lagrangians then take  the form:
\bea
 \mathcal{L}_{Z,\pm}&=&\frac{g}{c_{\lw}}\left\{ \frac{}{}g_{\lz\lf^+}\overline{\f^+}\gamma^\mu \f^++g_{\lz\lf^-}\overline{\f^-}\gamma^\mu \f^- +g^\ell_{L}\overline{\ell_{L}}\gamma^\mu\ell_{L} + g^\ell_{R}\overline{\ell_{R}}\gamma^\mu\ell_{R} \right.\nonumber\\
& & \left.\qquad\qquad-  \ \frac{1}{\sqrt{3}}\left[\overline{\ell_L^c}\,\mathcal{K}_{\s}\, \gamma^\mu \,\f^+_R+ \overline{\f^+_R}\,\gamma^\mu\,\mathcal{K}_{\s}^\dagger\,\ell_L^c\right]\right\}Z_\mu,\\
\mathcal{L}_{Z,0}&=&\frac{g}{c_{\lw}}\left\{ \frac{}{}g_{\lz\lf^0}\overline{\f^0}\gamma^\mu \f^0+g_{\nu}\overline{\nu_L}\gamma^\mu \nu_L - \, (\overline{\nu_L^c}\,U^T_{\nu}\,\mathcal{K}_{\s}\,\gamma^\mu \,\f^0_R+ \overline{\f^0_R} \,\gamma^\mu\,\mathcal{K}_{\s}^\dagger\,U^*_{\nu}\,\nu_L^c)\right\}Z_\mu,\nonumber
\eea
while the interaction Lagrangian for the $W$ boson becomes
\bea
\mathcal{L}_{W}&=& g\left\{\sqrt{\frac{3}{2}}\,\overline{\f^{+}}\gamma^\mu\f^0+\sqrt{2}\,\overline{\f^{0}}\gamma^\mu \f^-+\sqrt{\frac{3}{2}}\,\overline{\f^-}\gamma^\mu \f^{--}+\frac{1}{\sqrt{2}}\overline{\nu_L}\,U^\dagger_\nu\gamma^\mu \ell_L\right\}W^+_\mu\nonumber\\
& &+g\left\{-\sqrt{\frac{3}{2}}\,\overline{\ell_L}\,\mathcal{K}_{\Delta}^*\,\gamma^\mu\,\f_L^{--}-\frac{1}{\sqrt{2}}\overline{\f_L^0}\,\mathcal{K}_{\Delta}^T\,\gamma^\mu\,\ell_L-\frac{1}{\sqrt{2}}\,\overline{\nu_L}\,U_\nu^\dagger\,\mathcal{K}_{\Delta}^*\,\gamma^\mu\,\f^-_L\right.\nonumber\\
& &\left.-\sqrt{2}\,\overline{\ell^c_L}\,\mathcal{K}_\s\,\gamma^\mu\,\f_R^0-\frac{4}{\sqrt{6}}\,\overline{\f^+_R}\,\gamma^\mu\,\mathcal{K}_\s^\dagger\,U_\nu^*\,\nu_L^c-\sqrt{2}\,\overline{\nu_L^c}\,U_\nu^T\,\mathcal{K}_\s\,\gamma^\mu\,\f^-_R\right.\nonumber\\
& &\left.-\sqrt{\frac{3}{2}}\,\overline{\f^+_L}\,\gamma^\mu\,\mathcal{K}_{\Delta} ^T\,U_\nu\,\nu_L\right\}W_\mu^+\ +\ \mathrm{H.c.}\label{W_int_mass_basis}
\eea

To determine the decay properties of the exotics  one must also account for the mass-mixing with the Higgs boson. As mentioned already,  due to the relation $\langle S^0\rangle \ll \langle \Delta^0\rangle$ one can neglect the mixing of $S^0$ with $H^0$ relative to the (already small) mixing between $\Delta^0$ and $H^0$. Writing the Yukawa Lagrangian for $\Delta$ in terms of the mass eigenstates, Eq.~\eqref{delta_yukawa_lagrangian} gives
\bea
\mathcal{L}& \supset& -\lambda_\Delta \bar{L}\mathcal{F}_R\Delta\ \simeq\ -\frac{\theta_0}{\sqrt{3}} \,\lambda_\Delta\, h\, (\overline{\nu_L}\, U_\nu^\dagger \,\f^0_R +\overline{e_L}\,\f^-_R) +\ldots \label{lagrange_fermion_sm_higgs}
\eea
where $h$ denotes the SM-like neutral scalar and $\theta_0$ is the mixing angle. These vertices open-up additional decay channels for the exotic fermions $\f^0$ and $\f^-$ to final-states containing the SM-like scalar. With this information we can proceed to study the decay properties of the exotics.
%%%%%%%%%%%%%%%%%%%%%%%%%%%%%%%%%%%%%%%%%%%
\section{Exotic Fermion Decays\label{sec:exotic_fermion_decay}}
There are two classes of decays available to the exotic fermions.  Decays into purely SM final-states, for example $\f\rightarrow SM+SM'$, are sensitive to the mass-mixing and depend on the Yukawa coupling matrices. Decays of the heavier exotics into the lighter ones, $\f^Q\rightarrow\f^{Q'}+SM$, do not depend on the parameters in the Yukawa Lagrangian. We consider both types of decays in what follows but first we must discuss the mass-splitting between the components of $\f$.

\begin{figure}[ttt]
\begin{center}
        \includegraphics[width = 0.6\textwidth]{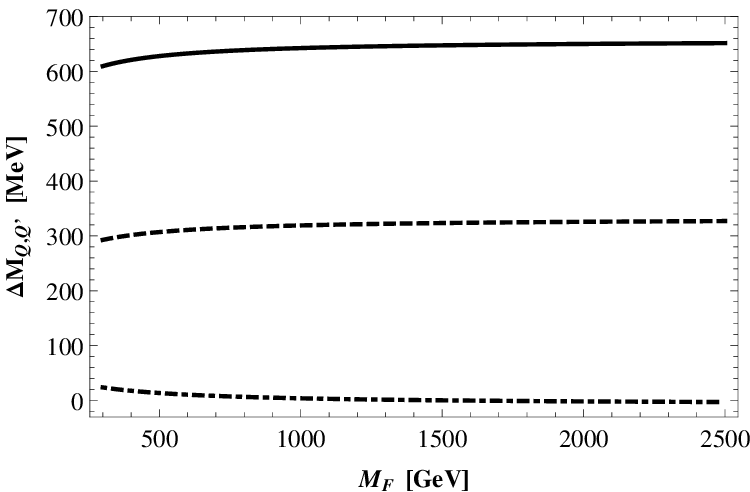}
\end{center}
\caption{Radiative mass-splitting due to loops with SM gauge bosons: $\Delta M_{Q,Q'}=M_{\f^Q}-M_{\f^{Q'}}$. The solid (dashed, dot-dashed) line corresponds to $\Delta M_{-2,-1}$ ($\Delta M_{-1,0}$, $\Delta M_{+1,0}$). For $M_{\f}\lesssim1.5$~TeV the mass ordering is $M_{\f^{--}}>M_{\f^{-}}>M_{\f^{+}}>M_{\f^0}$.}\label{fig:mass_splitting}
\end{figure}

Neglecting the tiny mixing with charged SM leptons, the components of $\f$ are degenerate at tree-level. This mass degeneracy is lifted by radiative corrections, with the dominant effect coming from loops with SM gauge bosons for $M_\f$ in the range of interest for the LHC.\footnote{Smaller $M_\f$ likely implies some Yukawa suppression of the the neutrino masses, which in turn suppresses Yukawa-induced loop-corrections to the exotic fermion masses.} At the one-loop level, the SM gauge bosons give the following calculable mass-splitting between the charged components $\f^Q$ and $\f^{Q'}$~\cite{Cirelli:2005uq},
\bea
\Delta M_{Q,Q'}&\equiv&M_Q-M_{Q'}\nonumber\\
&=& \frac{\alpha M_{\f}}{4\pi}\left\{(Q^2-Q'^2)f(r_Z) +s_{\lw}^{-2} (Q-Q')(Q+Q'-Y)[f(r_W)-f(r_Z)]\right\},\nonumber
\eea
where  $r_{Z,W}=M_{Z,W}/M_{\f}$ and the hypercharge value is $Y=-1$ in the present case.  For fermionic multiplets the function $f(r)$ has the form
\bea
f(r)=\frac{r}{2}\,\left\{2r^3\ln r-2r+(r^2-4)^{1/2}(r^2+2)\ln\left[\frac{r^2-2-r\sqrt{r^2-4}}{2}\right]\right\}.
\eea
 We plot this loop-induced mass-splitting $\Delta M_{Q,Q'}$ in Figure~\ref{fig:mass_splitting}. For the range of interest at the LHC, namely $M_{\f}\lesssim1$~TeV, the mass ordering is:
\bea
M_{\f^{--}}>M_{\f^{-}}>M_{\f^{+}}>M_{\f^0}.
\eea
As an  example, for $M_{\f}=300$~GeV the splittings are
\bea
M_{\f^{--}}-M_{\f^{-}}&\approx& 600~\mathrm{MeV}\,,\nonumber\\
 M_{\f^{-}}~-M_{\f^{0}}&\approx& 300~\mathrm{MeV}\,,\\
 M_{\f^{+}}~-M_{\f^{0}}&\approx& 20~\mathrm{MeV}.\nonumber
\eea
The plot shows that increasing $M_{\f}$ does not affect the ordering of $\f^{--}$ and $\f^{-}$, but the ordering of $\f^{+}$ and $\f^0$ can change to $M_{\f^0}-M_{\f^{+}}=\mathcal{O}(1~\mathrm{MeV})>0$ for $M_{\f}\gtrsim2$~TeV. This is seen in the figure, where $M_{+1,0}$ becomes negative for larger values of $M_\f$.  However, for $M_\f\lesssim1$~TeV the charged components are always heaviest. Observe that all splittings are below the mass of the $\rho(770)$ resonance.

%---------------------------------------------------------
%---------------------------------------------------------
%%%%%%%%%%%%%%%%%%%%%%%%%%%%%%%%%%%%%%%%%%%
\subsection{Decays to Standard Model Final-States\label{susec:decays_to_sm}}
We first discuss the decays $\f\rightarrow SM$, which are dominated by two-body final states. These decays arise in two ways. Firstly, the mass-mixing with leptons induces a coupling between $\f$ and SM leptons via the charged and neutral currents. Secondly, the Yukawa Lagrangian allows decays to SM final-states due to the mass-mixing between the SM Higgs and the exotic scalars. However, as discussed in Section~\ref{sec:new_seesaw}, one expects $\langle S^0\rangle\ll \langle \Delta^0\rangle$, given that the former is induced by the latter. Consequently the mixing between $S^0$ and $H^0$ is smaller than the (already small) mixing between $\Delta^0$ and $H^0$, and can be neglected. Decays induced by scalar mass-mixing therefore proceed predominantly through the couplings $\lambda_\Delta$, or equivalently the matrices $\mathcal{K}_\Delta$. Gauge-mediated decays can proceed through both coupling matrices $\mathcal{K}_\Delta$ and $\mathcal{K}_\s$, but due the relation $\langle S^0\rangle\ll \langle \Delta^0\rangle$ one expects the $\mathcal{K}_\Delta$-dependent pieces to dominate. These features play a role in the following.

Consider the neutral fermion $\f^0$. It has two-body SM decays containing a final state $W$ boson, with the decay widths
\bea
\Gamma(\f^0\rightarrow W^+\ell^- )&=& \frac{\alpha}{8s_{\lw}^2}\frac{1}{2}|\mathcal{K}^\ell_{\Delta}|^2 \frac{M_{\f}^3}{M_W^2}\left(1-\frac{M_W^2}{M_{\f}^2}\right)^{2} \left( 1+\frac{2M_W^2}{M_{\f}^2}\right)\,,\nonumber\\
\Gamma(\f^0\rightarrow W^-\ell^+ )&=& \frac{\alpha}{8s_{\lw}^2}2|\mathcal{K}^\ell_\s|^2 \frac{M_{\f}^3}{M_W^2}\left(1-\frac{M_W^2}{M_{\f}^2}\right)^{2} \left( 1+\frac{2M_W^2}{M_{\f}^2}\right)\,.
\eea
Neutral-current  decays are also possible:
\bea
\sum_{i=1}^3\Gamma(\f^0\rightarrow Z \nu_i)&=&  \sum_{\ell}\frac{\alpha}{8s_{\lw}^2c_{\lw}^2}|\mathcal{K}^\ell_\s|^2\frac{M_{\f}^3}{M_Z^2}\left(1-\frac{M_Z^2}{M_{\f}^2}\right)^{2} \left( 1+\frac{2M_Z^2}{M_{\f}^2}\right)\,.\label{neutralF_Z_decays}
\eea
To leading order, decays to the SM-like neutral scalar $h$ have the width 
\bea
\sum_i\Gamma(\f^0\rightarrow \nu_i\, h)
&=&\sum_{\ell}\frac{\theta_0^2}{96\pi}\,|\lambda_{\Delta,\ell}|^2\,M_{\f}\,\left(1-\frac{M_h^2}{M_{\f}^2}\right)^{2} \,\nonumber\\
&=& \sum_{\ell}\frac{1}{16\pi}\,|\mathcal{K}^\ell_\Delta|^2\,\frac{M_{\f}^3}{\langle H^0\rangle^2}\,\left(1-\frac{M_h^2}{M_{\f}^2}\right)^{2} \,.
\eea
 
Due to the $\mathcal{K}_\s$-dependence in Eq.~\eqref{neutralF_Z_decays}, decays to $Z$ bosons and $W^-\ell^+$ final-states are suppressed relative to the other channels. The dominant decays are $\sum_i\f^0\rightarrow \nu_ih$ and $\f^0\rightarrow W^+\ell^-$, with the $W^+\ell^-$ final state preferred over the $W^-\ell^+$ mode by a factor of
\bea
\frac{\Gamma(\f^0\rightarrow W^+\ell^- )}{\Gamma(\f^0\rightarrow W^-\ell^+)}&\approx&\frac{1}{4} \frac{|\mathcal{K}^\ell_{\Delta}|^2}{|\mathcal{K}^\ell_{\s}|^2} \ \gg\ 1\quad\quad \mathrm{for}\quad|\mathcal{K}_\Delta|\gg|\mathcal{K}_\s|\,.\label{eq_fneutral_width_ratio}
\eea
The relative branching fractions for electrons, muons and tauons are somewhat sensitive to the neutrino mass hierarchy, as will be discussed in the next section. However, this dependence drops out after summing over flavors. Note that,  because $\f^0$ is the lightest exotic in the parameter space of interest for the LHC, no decays of the type $\f^0\rightarrow \f'+SM$ are possible.

\begin{figure}[ttt]
\begin{center}
        \includegraphics[width = 0.6\textwidth]{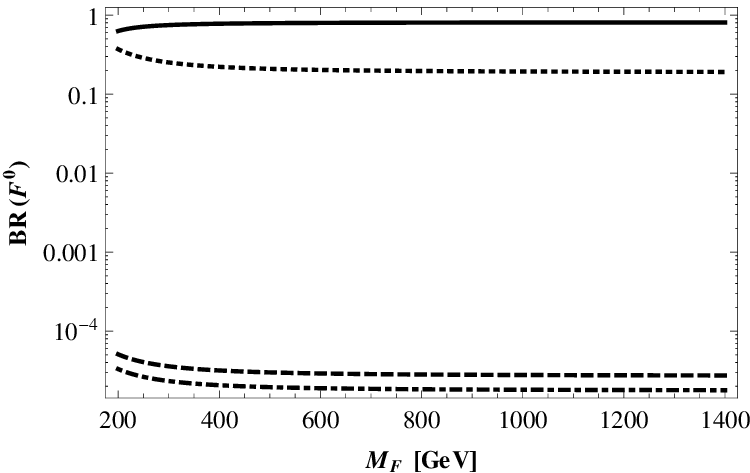}
\end{center}
\caption{Dominant branching fractions for the neutral fermion $\f^0$. Solid line =  $\sum_\nu \Gamma(\f^0\rightarrow\nu h)$, dotted  = $\sum_\ell\ell^-W^+$, dashed  = $\sum_\ell\ell^+W^-$, dot-dashed = $\sum_iZ\nu_i$. }\label{fig:Fneutral_BF_IH}
\end{figure}

In Figure~\ref{fig:Fneutral_BF_IH}  we plot the branching fractions for $\f^0$ as a function of $M_\f$, with  $|\mathcal{K}_\Delta|=10^{-6}\sqrt{\frac{2500}{(M_\f/\mathrm{GeV})}}\simeq 150\times|\mathcal{K}_\s|$. For $|\mathcal{K}_\Delta|\gg |\mathcal{K}_\s|$ the branching fractions are not very sensitive to the overall scale of $|\mathcal{K}_\Delta|$. The dominant modes are the $W^+\ell^-$ and $\nu h$ channels, as expected; the $\nu Z$   and $W^-\ell^+$ modes are greatly suppressed by the small factor $|\mathcal{K}_\s|^2/|\mathcal{K}_\Delta|^2\ll1$.  The key features observed in Figure~\ref{fig:Fneutral_BF_IH} persist as one increases the hierarchy between $|\mathcal{K}_\Delta|$ and $|\mathcal{K}_\s|$.\footnote{A larger hierarchy can  arise naturally given that $\langle S^0\rangle\ne0$ is induced by $\langle \Delta^0\rangle\ne0$ and that $\langle S^0\rangle\propto\lambda$.} Values of $|\mathcal{K}_\Delta|\gg 150|\mathcal{K}_\s|$ tend to further suppress the already-small branching fraction for $\sum_\nu\f^0\rightarrow \nu Z$ and $\f^0\rightarrow W^-\ell^+$, while the similarity of the widths for the $\nu h$ and $\ell^-W^+$ channels persists.  Figure~\ref{fig:Fneutral_BF_IH} therefore  provides a relatively robust representation of the dominant decay channels for $\f^0$.

%---------------------------------------------------------
%---------------------------------------------------------

Turning now to the positively charged fermion $\f^+$, it has two-body decays to SM final states containing neutrinos:
\bea
& &\sum_{i=1}^3\Gamma(\f^+\rightarrow W^+\nu_i)\nonumber\\
&=&\sum_{\ell}\frac{\alpha}{8s_{\lw}^2}\left[\frac{3}{2}|\mathcal{K}^\ell_\Delta|^2+\frac{8}{3}|\mathcal{K}^\ell_{\s}|^2\right] \frac{M_{\f}^3}{M_W^2}\left(1-\frac{M_W^2}{M_{\f}^2}\right)^{2} \left( 1+\frac{2M_W^2}{M_{\f}^2}\right)\,,\nonumber\\
\eea
 and  charged leptons:
\bea
\Gamma(\f^+\rightarrow Z \ell^+)&=&  \frac{\alpha}{8s_{\lw}^2c_{\lw}^2}\frac{1}{3}|\mathcal{K}^\ell_{\s}|^2\frac{M_{\f}^3}{M_Z^2}\left(1-\frac{M_Z^2}{M_{\f}^2}\right)^{2} \left( 1+\frac{2M_Z^2}{M_{\f}^2}\right)\,.
\eea
There is also a decay $\f^+\rightarrow \f^0+SM$, however, we will see in Section~\ref{susec:decays_to_light_F} that the corresponding width is negligible compared with the above. Therefore the $W^+\nu$ mode is dominant, with a branching fraction of  $\sim100\%$, given the $\mathcal{K}_\s$-dependence of the $Z\ell^+$ mode. This feature is generic in the present model; $\f^+$ is expected to decay dominantly as $\f^+\rightarrow W^+\nu$, giving signals with large amounts of missing energy.

The negatively-charged fermions $\f^-$ and $\f^{--}$ can decay to two-body SM final-states and to states with lighter exotic fermions (see below). The fermion $\f^-$ has two-body  decays to a $W$ boson and a neutrino:
\bea
& &\sum_{i=1}^3\Gamma(\f^-\rightarrow W^-\nu_i)\nonumber\\
&=&\sum_{\ell}\frac{\alpha}{8s_{\lw}^2}(2|\mathcal{K}^\ell_\s|^2+\frac{1}{2}|\mathcal{K}^\ell_{\Delta}|^2)\frac{M_{\f}^3}{M_W^2}\left(1-\frac{M_W^2}{M_{\f}^2}\right)^{2} \left( 1+\frac{2M_W^2}{M_{\f}^2}\right)\,.
\eea
The decay $\f^-\rightarrow Z\ell^-$  does not happen at leading order  due to the cancellation in the neutral-current interaction Lagrangian, resulting from the equality $g_L^\ell=g_{\lz\lf^-}$. The width for  $\f^-\rightarrow \ell^-+h$ has the leading-order value 
\bea
\Gamma(\f^-\rightarrow \ell^-\, h)
&=&\frac{\theta_0^2}{96\pi}\,|\lambda_{\Delta,\ell}|^2\,M_{\f}\,\left(1-\frac{M_h^2}{M_{\f}^2}\right)^{2} \,\nonumber\\
&=& \frac{1}{16\pi}\,|\mathcal{K}^\ell_\Delta|^2\,\frac{M_{\f}^3}{\langle H^0\rangle^2}\,\left(1-\frac{M_h^2}{M_{\f}^2}\right)^{2} \,.
\eea
Observe that for $|\mathcal{K}_\Delta|\gg|\mathcal{K}_\s|$ one expects 
\bea
\sum_{i}\Gamma(\f^-\rightarrow W^-\nu_i)&\sim& \sum_\ell\Gamma(\f^-\rightarrow \ell^-\, h)\,,
\eea
which is contrary to $\f^+$, for which the decay $\f^+\rightarrow W^+\nu_i$ dominates the charged-lepton modes.

%---------------------------------------------------------
\begin{figure}[ttt]
\begin{center}
        \includegraphics[width = 0.6\textwidth]{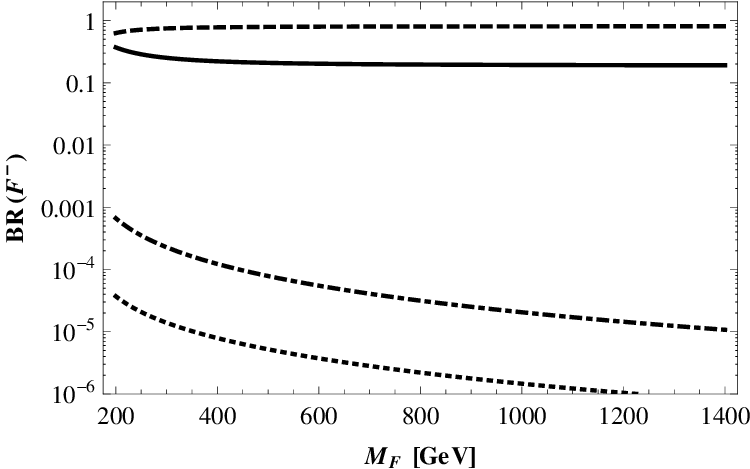}
\end{center}
\caption{ Branching fractions for the negatively-charged fermion $\f^{-}$. Solid line = $\sum_\nu\Gamma(\f^-\rightarrow W^-\nu)$, dashed  = $\sum_\ell \ell^-h$, dotdashed  = $ \f^0\pi^-$, dotted  = $\sum_{\ell=e,\mu} \f^0 \ell^-\nu$.}\label{fig:Fminus_BF_small_Ks_IH}
\end{figure}
%---------------------------------------------------------

The doubly-charged fermion $\f^{--}$ has a single two-body decay mode to SM states. It proceeds via the charged current and has the width:
\bea
\Gamma(\f^{--}\rightarrow W^-\ell^- )&=& \frac{\alpha}{8s_{\lw}^2}\frac{3}{2}|\mathcal{K}^\ell_{\Delta}|^2 \frac{M_{\f}^3}{M_W^2}\left(1-\frac{M_W^2}{M_{\f}^2}\right)^{2} \left( 1+\frac{2M_W^2}{M_{\f}^2}\right)\,.
\eea
%%%%%%%%%%%%%%%%%%%%%%%%%%%%%%%%%%%%%%%%%%%
\subsection{Decays to Light Exotic Fermions\label{susec:decays_to_light_F}}

In addition to the two-body decays $\f\rightarrow\mathrm{SM+SM'}$, the heavier fermions can decay to the lighter ones, $\f\rightarrow\f'+\mathrm{SM}$. The widths for these decays do not depend on any unknown parameters; they are determined by the mass-splitting between $\f$ and $\f'$, which depends only weakly on the mass $M_\f$, as seen in Figure~\ref{fig:mass_splitting}. For the region for interest at the LHC, the widths are essentially independent of $M_\f$  and are determined by the quantum numbers of $\f$.

%---------------------------------------------------------
\begin{figure}[ttt]
\begin{center}
        \includegraphics[width = 0.6\textwidth]{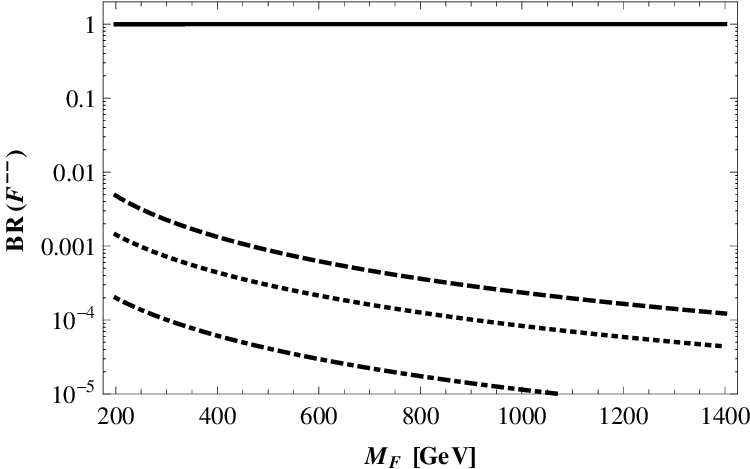}
\end{center}
\caption{Branching fractions for the doubly-charged fermion $\f^{--}$. Solid line = $\sum_\ell\Gamma(\f^{--}\rightarrow W^-\ell^-)$, dashed  = $ \f^-\pi^-$,  dotted  = $\sum_{\ell=\mu,e}\f^-\ell^-\nu$, dotdashed =  $ \f^-K^-$.}\label{fig:Fdoublyminus_BF_small_Ks_IH}
\end{figure}
%---------------------------------------------------------

The negatively-charged fermions can decay to single-pion final states, with partial decay widths given by
\bea
\Gamma(\f^{Q}\rightarrow \f^{Q+1}\,\pi^-)&=&\frac{2g_{\lw\lf}^2|V_{ud}|^2}{\pi}\,G_F^2\,f_\pi^2\,(\Delta M)^3\left(1-\frac{m_\pi^2}{(\Delta M)^2}\right)^{1/2}\,.
\eea
Here $Q<0$, and the couplings are  $g_{\lw\lf}^2= \{3/2,\,2\}$ for $Q=\{-2,-1\}$, which can be read off Eq.~\eqref{W_int_mass_basis}.  The mass-splitting between the  fermions is denoted by $\Delta M$. For $\f^{--}$ the kaon mode $\f^{--}\rightarrow \f^-K^-$ is also kinematically accessible, but the width is suppressed relative to the pion mode by the small CKM factor $|V_{us}|^2\ll|V_{ud}|^2$. The three-body decay $\f^{--}\rightarrow W^-\pi^-\pi^0$ is also kinematically allowed. However, the mass-splitting  between $\f^{--}$ and $\f^-$ is $\Delta M\approx 600$~MeV, which is below the mass of the $\rho(770)$ resonance. The width is therefore suppressed by the three-body phase space and the small mass differences. Three-body decays to kinematically accessible final-state leptons are also available for the negatively charged fermions:
\bea
\Gamma(\f^Q\rightarrow \f^{Q+1}\,\ell^-\nu)&=&\frac{2g_{\lw\lf}^2}{15\pi^3}\,G_F^2\,(\Delta M)^5\left(1-\frac{m_\ell^2}{(\Delta M)^2}\right)^{1/2}\,P(m_\ell/\Delta M),
\eea
where 
\bea
P(X)= 1-\frac{9}{2}X^2-4X^4+\frac{15X^4}{2\sqrt{1-X^2}}\tanh^{-1}\sqrt{1-X^2}.
\eea
These widths are also suppressed by the small mass difference, particularly for $\f^-$.

The positively-charged fermion cannot kinematically access final-states containing tauons, pions or muons, due to the small $\mathcal{O}(10~\mathrm{MeV})$ mass-splitting between $\f^+$ and $\f^0$. The only kinematically-accessible decay of the type $\f^+\rightarrow\f^0+\mathrm{SM}$ is to a three-body final state with a positron: 
\bea
\Gamma(\f^+\rightarrow \f^{0}\,e^+\nu)&=&\frac{3}{15\pi^3}\,G_F^2\,(\Delta M)^5\left(1-\frac{m_e^2}{(\Delta M)^2}\right)^{1/2}.
\eea
However, this decay is highly suppressed due to the tiny mass-splitting between $\f^+$ and $\f^0$, and can be ignored.

With the above results we can  plot the branching fractions for $\f^-$ and $\f^{--}$, as shown in Figures~\ref{fig:Fminus_BF_small_Ks_IH} and \ref{fig:Fdoublyminus_BF_small_Ks_IH}, respectively. The following features can be noted. In both cases the modes $\f\rightarrow\f'+SM$ are subdominant to the $\f\rightarrow SM+SM'$ modes. For example, one has
\bea
\sum_{\ell}BR(\f^{--}\rightarrow W^-\ell^-)&>&  99.9\%\quad\quad\mathrm{for}\quad M_\f=400~\mathrm{GeV},
\eea
and $BR(\f^-\rightarrow SM+SM')$ is even larger. The relative size of the  widths for the $\f\rightarrow SM+SM'$ modes and the $\f\rightarrow \f'+SM$ modes depends on $|\mathcal{K}_\Delta|$. However,  given that $|\mathcal{K}_\Delta|\gg|\mathcal{K}_\s|$ is expected and that $m_\nu\sim \mathcal{K}_\Delta M_\f \mathcal{K}_\s$, the demand of $m_\nu\sim0.1$~eV  means that $|\mathcal{K}_\Delta|$ is not expected to be small enough to change the relation $\Gamma(\f\rightarrow SM+SM')>\Gamma(\f\rightarrow \f'+SM)$ for $M_\f\lesssim$~TeV. Increasing the hierarchy between $|\mathcal{K}_\Delta|$ and $|\mathcal{K}_\s|$ increases this inequality so that both $\f^-$ and $\f^{--}$ are expected to  decay  dominantly as $\f\rightarrow SM+SM'$, as in Figures~\ref{fig:Fminus_BF_small_Ks_IH} and \ref{fig:Fdoublyminus_BF_small_Ks_IH}. This means that $\f^-$ always has a sizable width to two-body finals states with charged leptons.

%%%%%%%%%%%%%%%%%%%%%%%%%%%%%%%%%%%%%%%%%%%%%%%%%%%%%%%
\section{Collider Signals\label{sec:signals}}

We have seen that the dominant decay modes for the exotic fermions are of the type
\bea
\f&\rightarrow& \mathrm{SM~lepton}\ +\ \mathrm{SM~boson},
\eea where the boson can be a $W$, $Z$ or $h$. Collider production of fermion pairs thus leads to events with pairs of SM leptons and bosons. We list the events with two charged-leptons in Table~\ref{all_charged_lepton_signals}, along with the branching fractions, for $M_\f=300$~GeV. The results in the table have limited  (no) sensitivity to the neutrino mass ordering (mixing phases), though we shall see later that the flavor content of the charged leptons is sensitive to both the mass-hierarchy and the phases. We include modes like $\f^+\rightarrow \ell^+Z$ in the table, despite their tiny branching fractions, to emphasize the absence of such events. The table uses $|\mathcal{K}_\Delta|\simeq150|\mathcal{K}_\s|$ and we  checked numerically that increasing the hierarchy between $|\mathcal{K}_\Delta|$ and $|\mathcal{K}_\s|$ barely changes the branching fractions. 

 The events listed in Table~\ref{all_charged_lepton_signals}  are in direct correspondence with those given in Ref.~\cite{delAguila:2008cj} for the Type-III seesaw.  The table shows that a number of dilepton events are possible.  The subsequent decay of final-state bosons produces various three- and four-lepton final states, giving a host of multi-lepton signatures whose discovery prospects are largely detailed in Ref.~\cite{delAguila:2008cj}.  Note that like-sign dilepton events, which  break lepton-number symmetry, are possible, but have highly suppressed branching fractions.  This is due to the dependence on the small parameter $|\mathcal{K}_\s|$. The branching fractions are much smaller  than those quoted  for the Type-III seesaw~\cite{delAguila:2008cj}, and are not expected to be accessible at the LHC.\footnote{We emphasize that we employ the hierarchy $|\mathcal{K}_\Delta|\gg|\mathcal{K}_\s|$, which is the origin of the suppressed branching fractions for like-sign lepton events. In principle one could consider a hierarchy in the Yukawa couplings to overcome the hierarchy in VEVs and give $|\mathcal{K}_\Delta|\sim|\mathcal{K}_\s|$; this would allow one to increase the branching fraction for like-sign lepton events but seems contrary to the spirit of the model.} It is clear from the table that charged lepton events are only have sizable branching-fractions  when they proceed through the coupling $|\mathcal{K}_\Delta|$. Thus, production of the pairs $\f^{--}\overline{\f^{--}}$, $\f^{--}\overline{\f^{-}}$, $\f^{0}\overline{\f^-}$, $\f^0\overline{\f^0}$, $\f^-\overline{\f^-}$, and the charge conjugate pairs, are most promising.

\begin{comment}
The like-sign dilepton events have the branching fractions 
\bea
BR(\f^0\overline{\f^0}\rightarrow \ell^{+}\ell^+ W^- W^-)&=& BR(\f^0\overline{\f^0}\rightarrow \ell^{-}\ell^- W^+ W^+)\ \simeq\ 5\%,\nonumber\\
BR(\f^-\overline{\f^0}\rightarrow \ell^{-}\ell^- W^+ h)&=& BR(\f^0\overline{\f^-}\rightarrow \ell^{+}\ell^+ W^-h)\ \simeq\ 8\%.
\eea
These values  are similar to those quoted  for the Type-III seesaw, which can be probed at the LHC~\cite{delAguila:2008cj}. Note, however, that the bosonic content of the like-sign dilepton events  differs for the two models; whereas the like-sign dileptons come with $WW$ or $Wh$ pairs in the present model, they are partnered with $ZW$ or $Wh$ pairs in the Type-III seesaw. 
\end{comment}
%%%%%%%%%%%%%%%%%%%%%%%%%%%%%%%%%%%%%%%%%%%%%%%%%%%%%%%

\begin{table}
\centering
\begin{tabular}{|c|c|c|c|c|c|}\hline
& & & & &\\
\ \ \ \ &
$\overline{\f^{--}}\rightarrow\ell^+W^+$ & $\overline{\f^{0}}\rightarrow\ell^+W^-$ &
$\overline{\f^{0}}\rightarrow\ell^-W^+$&$\overline{\f^{+}}\rightarrow\ell^-Z$&$\overline{\f^{-}}\rightarrow\ell^+h$\\
&(0.997)& (0.25)&($10^{-5}$) &($10^{-5}$) &(0.75)\\
\hline
%-----------------------------------------------%
& & & & &\\
$\f^{--}\rightarrow\ell^-W^-$& $\ell^-\ell^+W^-W^+$ &$-$&$-$&$-$ &$\ell^-\ell^+W^-h$\\ 
(0.997)&(0.99)&&&&(0.75)\\
\hline
%------------------------------------------------%
& & & & &\\
$\f^{0}\rightarrow\ell^-W^+$& $- $&$\ell^-\ell^+W^-W^+$ &$\ell^-\ell^-W^+W^+$&$\ell^-\ell^-W^+Z$ &$\ell^-\ell^+W^+h$\\ 
(0.25)& & (0.06)& ($10^{-5}$)& ($10^{-6}$)&(0.19)\\
\hline
%------------------------------------------------%
& & & & &\\
$\f^{0}\rightarrow\ell^+W^-$& $-$ &$\ell^+\ell^+W^-W^-$ &$\ell^-\ell^+W^-W^+$&$\ell^-\ell^+W^-Z$ &$\ell^+\ell^+W^-h$\\ 
($10^{-5}$)& & ($10^{-5}$)& ($10^{-9}$)& ($10^{-10}$)&($10^{-5}$)\\
\hline
%------------------------------------------------%
& & & & &\\
$\f^{+}\rightarrow\ell^+Z$& $-$ &$\ell^+\ell^+W^-Z$ &$\ell^-\ell^+W^+Z$&$\ell^-\ell^+ZZ$&$-$\\ 
($10^{-5}$)& & ($10^{-6}$)&($10^{-10}$) & ($10^{-10}$)&\\
\hline
%------------------------------------------------%
& & & & &\\
$\f^{-}\rightarrow\ell^-h$& $\ell^-\ell^+W^+h$ &$\ell^-\ell^+W^-h$ &$\ell^-\ell^-W^+h$&$-$&$\ell^-\ell^+hh$\\ 
(0.75)&(0.75)& (0.19)& ($10^{-5}$)& & ($0.56$)\\
\hline
%------------------------------------------------%
\end{tabular}
\caption{\label{all_charged_lepton_signals} Events due to exotic fermion decays  to charged-leptons, $\ell=e,\,\mu,\,\tau$.  Branching fractions are shown for $M_\f=300$~GeV. Results are basically the same for both  the normal and inverted hierarchies,  and hold for arbitrary values of the neutrino phases.}
\end{table}
%%%%%%%%%%%%%%%%%%%%%%%%%%%%%%%%%%%%%%%%%%%%%%%%%%%%%%%

In addition to the events shown in Table~\ref{all_charged_lepton_signals} there are various events containing neutrinos. Events with one charged lepton are possible, like $\ell^-W^-W^+\nu$, and there are also events with no charged leptons; for example, pair production of $\f^+\overline{\f^+}$ gives $W^+W^-\nu\nu$ final states with a branching fraction of $\sim100\%$.  Both classes of events give signals with large amounts of missing energy.

 The events in Table~\ref{all_charged_lepton_signals}  can  be compared with the corresponding table  in Ref.~\cite{Kumericki:2012bh}, where   similar signals are listed for the seesaw/radiative model $(b)$, which employs the fermion $\f\sim(1,5,0)$ (see Table~2 in Ref.~\cite{Kumericki:2012bh}).  Similar to the Type-III seesaw, model $(b)$ predicts $\ell^\pm\ell^\pm W^\mp Z$ events but not $\ell^\pm\ell^\pm W^\mp W^\mp$ events~\cite{Kumericki:2012bh}, which is opposite to the present model. The difference arises because the neutral beyond-SM fermion has zero hypercharge in both model $(b)$ and the Type-III seesaw;  production of neutral fermion pairs, which would otherwise give the lepton-number violating event $\ell^\pm \ell^\pm W^\mp W^\mp$,  is not available in those models.\footnote{More accurately, it can only proceed via Lagrangian terms containing two insertions of the small mass-mixing parameters and is therefore highly suppressed.}  The present model allows neutral-fermion pair-production and thus predicts $\ell^\pm \ell^\pm W^\mp W^\mp$ events, though the branching fraction is highly suppressed due to the relation $|\mathcal{K}_\s|\ll|\mathcal{K}_\Delta|$. This provides a key way to discriminate the present model from the Type-III seesaw and model $(b)$. The observation of like-sign dilepton events would provide a clear preference for those models.

\begin{comment}
Though the  $\ell^\pm\ell^\pm W^\mp W^\mp$  state is not expected in either model $(b)$ or the Type-III seesaw, the actual experimental signal requires one to decay the $W$ bosons. If these decay leptonically one  obtains four-lepton events via the chain
\bea
pp\ \rightarrow\ \bar{\f^0}\f^0\ \rightarrow\ \ell^\pm\ell^\pm W^\mp W^\mp\ \rightarrow\ \ell^\pm\ell^\pm \ell^\mp\ell^\mp +\mathrm{missing\ energy}\,.
\eea
This four-lepton signal can  arise via other intermediate states in all three models. However, the kinematics for the lepton-number violating channel $\ell^\pm\ell^\pm W^\mp W^\mp$ are quite distinct; one expects two like-sign  hard leptons of a given charge and two like-sign softer-leptons with the opposite charge.  This differs from the analogous state in both model $(b)$ and the Type-III seesaw (and also from other channels in the present model), for which one expects two opposite-sign hard leptons and two opposite-sign softer leptons. 
\end{comment}

%---------------------------------------------------------
\begin{figure}[ttt]
\centering
\begin{tabular}{cc}
\epsfig{file=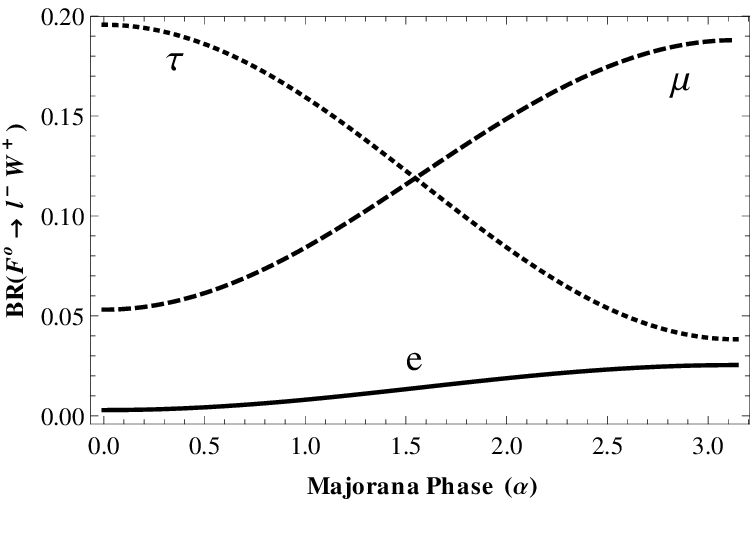,width=0.47\linewidth,clip=}&
\epsfig{file=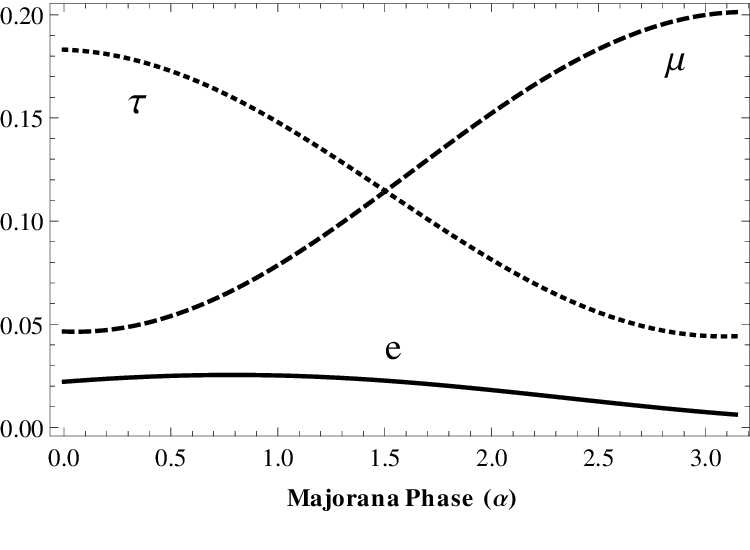,width=0.45\linewidth,clip=}
%(a)&(b)\\
\end{tabular}
\caption{Normal hierarchy: Branching fraction $BR(\f^0\rightarrow \ell^- W^+)$ as a function of the mixing phase $\alpha$. Left (right) plot is for $\delta=0$ ($\delta=3\pi/4$). }\label{BR_fneutral_delta_threePionfour_NH}
\end{figure}
%----------------------------------------------------

Note that $\ell^\pm\ell^\pm W^\mp W^\mp$ events can arise in model $(d)$, which employs the fermion $\f\sim(1,5,2)$ and contains a neutral fermion with nonzero hypercharge~\cite{Picek:2009is}. The branching fractions for such events have additional parameter dependence in model $(d)$ because the analogues of $\mathcal{K}_{\Delta}$ and $\mathcal{K}_{\s}$ are essentially independent parameters (the VEV appearing in one parameter is not induced by the VEV appearing in the other). This means that, in regions of parameter space where the couplings are of a similar magnitude, the branching fraction for $\ell^\pm\ell^\pm W^\mp W^\mp$ events can be non-negligible, allowing one to discriminate between models. There remains an alternative simple means of discriminating  the two models, as model $(d)$  contains a triply charged fermion that gives striking ``golden" decays of the type $\f^{+++}\rightarrow W^+W^+\ell^+$~\cite{Picek:2009is}. These  lead to a class of prominent signals~\cite{Picek:2009is} that will be absent  if model $(c)$ is realized in nature. The observation of $\ell^\pm\ell^\pm W^\mp W^\mp$ events in conjunction with golden decays would favor model $(d)$ over the present model.

%%%%%%%%%%%%%%%%%%%%%%%%%%%%%%%%%%%%%%%%%%%%%%%%%%%%%%%

\begin{table}
\centering
\begin{tabular}{|c|c|c|c|c|}\hline
& & & & \\
\ \ \ \ &
$\overline{\f^{--}}\rightarrow\ell^+W^+$ & $\overline{\f^{0}}\rightarrow\ell^+W^-$ &
$\overline{\f^{0}}\rightarrow\ell^-W^+$&$\overline{\f^{-}}\rightarrow\ell^+h$\\
&&&&\\
\hline
%-----------------------------------------------%
& & &  &\\
$\f^{--}\rightarrow\ell^-W^-$& $\ell^-\ell^+W^-W^+$ &$-$&$-$&$\ell^-\ell^+W^-h$\\ 
&(0.05) \,(0.60)&&&(0.04)\,(0.45)\\
\hline
%------------------------------------------------%
& & & & \\
$\f^{0}\rightarrow\ell^-W^+$&$-$  &$\ell^-\ell^+W^-W^+$ &$\ell^-\ell^-W^+W^+$ &$\ell^-\ell^+W^+h$\\ 
&& (0.003)\,(0.03)& ($10^{-5}$)\,($10^{-5}$)&(0.01)\,(0.11)\\
\hline
%------------------------------------------------%
& & & & \\
$\f^{0}\rightarrow\ell^+W^-$&$-$ &$\ell^+\ell^+W^-W^-$ &$\ell^-\ell^+W^-W^+$ &$\ell^+\ell^+W^-h$\\ 
&&  ($10^{-5}$)\,($10^{-5}$)& ($10^{-9}$)\,($10^{-9}$)&($10^{-5}$)\,($10^{-5}$)\\
\hline
%------------------------------------------------%
& & & & \\
$\f^{-}\rightarrow\ell^-h$& $\ell^-\ell^+W^+h$ &$\ell^-\ell^+W^-h$ &$\ell^-\ell^-W^+h$&$\ell^-\ell^+hh$\\ 
&(0.04)\,(0.45)& (0.01)\,(0.11)& $(10^{-5})\,(10^{-5})$& ($0.03$)\,(0.34)\\
\hline
%------------------------------------------------%
\end{tabular}
\caption{\label{NH_light_charged_lepton_signals} Normal hierarchy: Branching fractions for events containing  light charged-leptons  ($\ell=e,\,\mu$) for $M_\f=300$~GeV. The first (second) value corresponds to neutrino-mixing phases of $\alpha=\delta=0$ ($\alpha=5/2,\,\delta=3\pi/4$), which exemplify the pessimistic (optimistic) scenario for light-lepton events.}
\end{table}
%%%%%%%%%%%%%%%%%%%%%%%%%%%%%%%%%%%%%%%%%%%%%%%%%%%%%%%

\begin{figure}[ttt]
\centering
\begin{tabular}{cc}
\epsfig{file=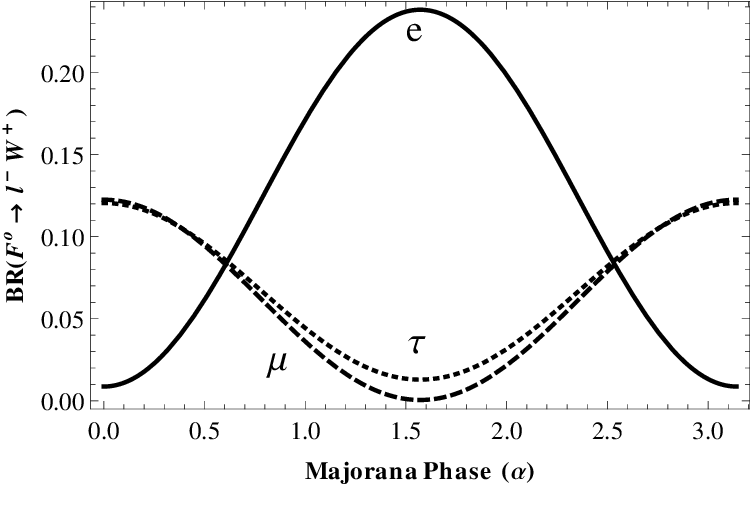,width=0.47\linewidth,clip=}&
\epsfig{file=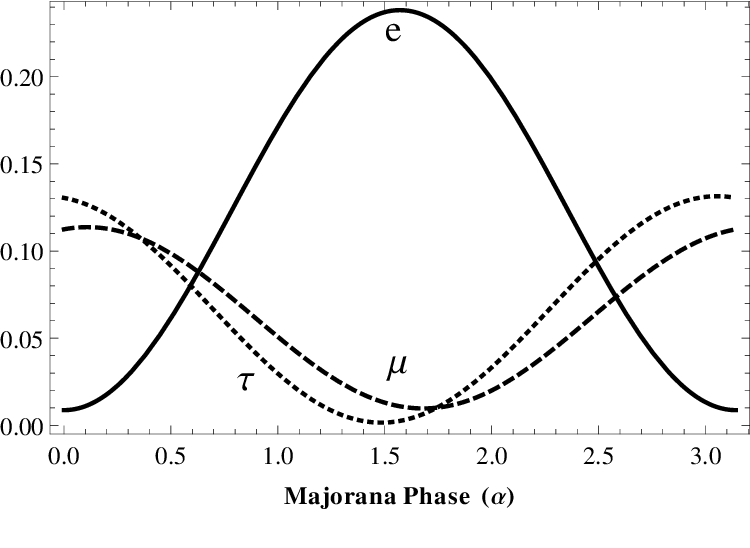,width=0.45\linewidth,clip=}
%(a)&(b)\\
\end{tabular}
\caption{Inverted  hierarchy: Branching fraction $BR(\f^0\rightarrow \ell^- W^+)$ as a function of the mixing phase $\alpha$. Left (right) plot is for $\delta=0$ ($\delta=3\pi/4$).}\label{BR_fneutral_delta_threePionfour_IH}
\end{figure}

 The branching fractions in Table~\ref{all_charged_lepton_signals} are largely insensitive to the mass ordering and the mixing phases $\alpha$ and $\delta$. However, it is important to differentiate between electron/muon events and tauon events as the former provide a much cleaner signal. We find that the branching fractions for distinct charged-leptons are sensitive to  both the mass ordering and the mixing phases. To demonstrate this dependence we plot $BR(\f^0\rightarrow\ell^-W^+)$ as a function of the phase $\alpha$, with $\delta=0$ and $\delta=3\pi/4$, for the case of a normal (inverted) mass hierarchy in  Figure~\ref{BR_fneutral_delta_threePionfour_NH} (\ref{BR_fneutral_delta_threePionfour_IH}). One observes significant differences between the two figures. For example, with a normal hierarchy the branching fraction for $\f^0\rightarrow e^-W^+$ is always small:\\
\bea
\frac{BR(\f^0\rightarrow e^-W^+)}{\sum_\ell BR(\f^0\rightarrow \ell^-W^+)}&\lesssim& 0.1,
\eea
while the relative size of the muon and tauon fractions varies with $\alpha$, as seen in Figure~\ref{BR_fneutral_delta_threePionfour_NH}. On the other hand, for an inverted hierarchy  there are large portions of parameter space in which the  electron modes are dominant, with values as large as
\bea
\frac{BR(\f^0\rightarrow e^-W^+)}{\sum_\ell BR(\f^0\rightarrow \ell^-W^+)}&\sim& 0.95,
\eea
in some regions. Furthermore the tauon fraction can be vanishingly small in regions of parameter space.\footnote{A similar phase-dependence of decay branching fractions occurs for a Type-II seesaw~\cite{Akeroyd:2007zv} but only when purely leptonic decays dominate~\cite{Kanemura:2013vxa}.} 

In Table~\ref{NH_light_charged_lepton_signals} we reproduce the charged-lepton events of Table~\ref{all_charged_lepton_signals} for the case of a normal hierarchy, restricting attention to light-lepton events ($\ell=e,\,\mu$). We select values of $\alpha$ and $\delta$ that demonstrate the range of  branching fractions expected. Comparison with Table~\ref{all_charged_lepton_signals}  shows that the branching fractions for light-lepton events can be less than, or on the order of, the branching fraction for tauon events. A sizable fraction of tauon events will of course reduce the signal for light-lepton searches.\footnote{The tauons will be very hard, however,  so the daughter lighter-leptons remain hard. Ref.~\cite{Aguilar-Saavedra:2013twa} argues that hard-daughter light-leptons can pass experimental selections with reasonable efficiency in the case of a Type-III seesaw. These arguments appear to hold in the present context.}

%---------------------------------------------------------
%----------------------------------------------------

 Table~\ref{IH_light_charged_lepton_signals}  lists the  branching fractions for light charged-lepton events in the case of an inverted hierarchy for $\alpha=\delta=0$ and $\alpha=3/2$, $\delta=3\pi/4$. These values demonstrate the range of branching fractions expected and can be compared with Table~\ref{NH_light_charged_lepton_signals}. The branching fractions are generally larger for the inverted hierarchy  and in many instances the discrepancy is significant. 

%%%%%%%%%%%%%%%%%%%%%%%%%%%%%%%%%%%%%%%%%%%%%%%%%%%%%%%

\begin{table}
\centering
\begin{tabular}{|c|c|c|c|c|}\hline
& & & & \\
\ \ \ \ &
$\overline{\f^{--}}\rightarrow\ell^+W^+$ & $\overline{\f^{0}}\rightarrow\ell^+W^-$ &
$\overline{\f^{0}}\rightarrow\ell^-W^+$&$\overline{\f^{-}}\rightarrow\ell^+h$\\
&&&&\\
\hline
%-----------------------------------------------%
& & &  &\\
$\f^{--}\rightarrow\ell^-W^-$& $\ell^-\ell^+W^-W^+$ &$-$&$-$&$\ell^-\ell^+W^-h$\\ 
&(0.27) \,(0.98)&&&(0.20)\,(0.74)\\
\hline
%------------------------------------------------%
& & & & \\
$\f^{0}\rightarrow\ell^-W^+$&$-$&$\ell^-\ell^+W^-W^+$ &$\ell^-\ell^-W^+W^+$ &$\ell^-\ell^+W^+h$\\ 
& & (0.02)\,(0.06)& ($10^{-6}$)\,($10^{-6}$)&(0.05)\,(0.19)\\
\hline
%------------------------------------------------%
& & & & \\
$\f^{0}\rightarrow\ell^+W^-$&$-$&$\ell^+\ell^+W^-W^-$ &$\ell^-\ell^+W^-W^+$ &$\ell^+\ell^+W^-h$\\ 
& &  ($10^{-6}$)\,($10^{-6}$)& ($10^{-9}$)\,($10^{-10}$)&($10^{-5}$)\,($10^{-5}$)\\
\hline
%------------------------------------------------%
& & & & \\
$\f^{-}\rightarrow\ell^-h$& $\ell^-\ell^+W^+h$ &$\ell^-\ell^+W^-h$ &$\ell^-\ell^-W^+h$&$\ell^-\ell^+hh$\\ 
&(0.20)\,(0.74)& (0.05)\,(0.19)& $(10^{-5})\,(10^{-5})$& ($0.15$)\,(0.55)\\
\hline
%------------------------------------------------%
\end{tabular}
\caption{\label{IH_light_charged_lepton_signals} Inverted hierarchy: Branching fractions for events containing light charged-leptons ($\ell=e,\,\mu$) for $M_\f=300$~GeV. The first (second) value corresponds to neutrino-mixing phases of $\alpha=\delta=0$ ($\alpha=3/2,\,\delta=3\pi/4$), which exemplify the pessimistic (optimistic) scenario for light-lepton events. }
\end{table}
%%%%%%%%%%%%%%%%%%%%%%%%%%%%%%%%%%%%%%%%%%%%%%%%%%%%%%%

\begin{figure}[ttt]
\centering
\begin{tabular}{cc}
\epsfig{file=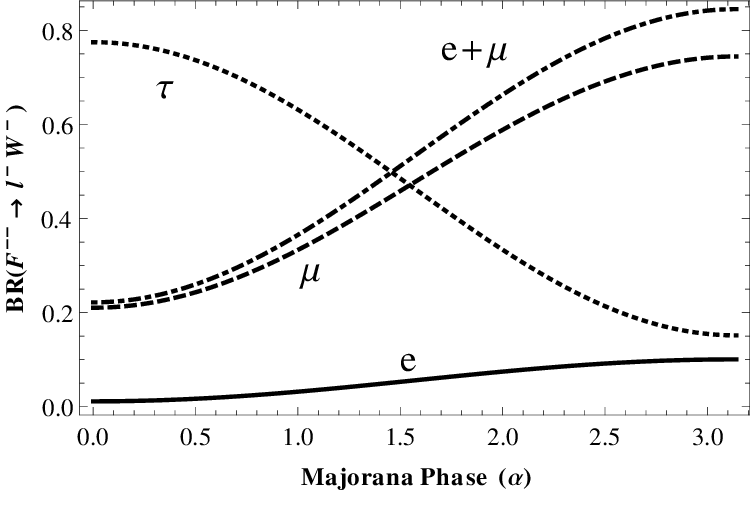,width=0.47\linewidth,clip=}&
\epsfig{file=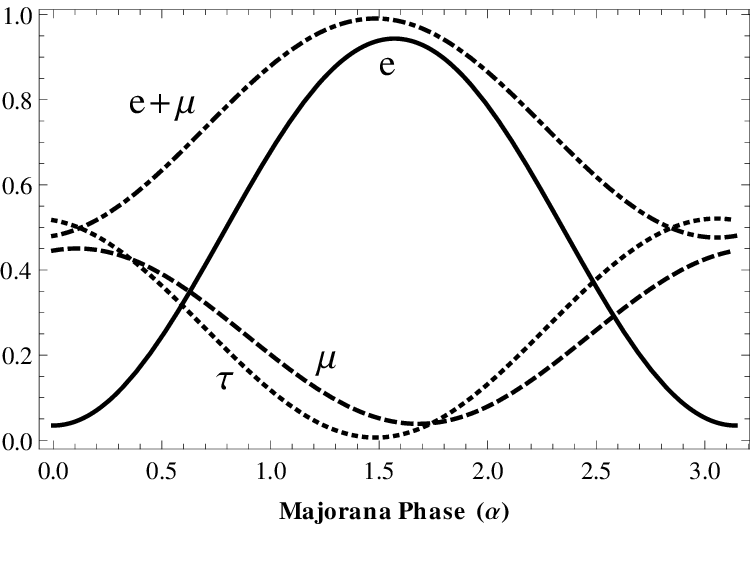,width=0.45\linewidth,clip=}
\end{tabular}
\caption{Leptonic branching fractions for the doubly-charged fermion, $BR(\f^{--}\rightarrow\ell^-W^-)$, as a function of the neutrino-mixing phase $\alpha$. The left (right) plot is for a normal (inverted) mass hierarchy with $\delta=0$ ($\delta=3\pi/4$), which is the more pessimistic (optimistic) scenario for discovery via light-lepton events.}\label{lepton_BR_fdoublyminus_delta_threePionfour}
\end{figure}

The discussion regarding the relative size of $BR(\f^0\rightarrow \ell^-W^+)$ for different leptons $\ell$  generalizes for the decays of the other components of $\f$. The doubly-charged fermion is of particular interest and we plot $BR(\f^{--}\rightarrow \ell^-W^-)$  as a function of the phase $\alpha$ in Figure~\ref{lepton_BR_fdoublyminus_delta_threePionfour}, with selected values of $\delta$ that demonstrate the range of branching fractions available. The $\alpha$-dependence of the relative branching fractions  is  similar to those found for $\f^0$, up to an overall scaling. In the figure the branching fraction for decays to light charged-leptons lies roughly in the range
\bea
BR(\f^{--}\rightarrow (e~\mathrm{or}~\mu)+W^-)\in\left\{
\begin{array}{cl}
[0.22,\,0.85]&\quad\mathrm{for\ a\ normal\ hierarchy}\\
\left[0.47,\,0.99\right]&\quad\mathrm{for\ an\ inverted\ hierarchy},
\end{array}\right.\label{eq:bf_range_f--}
\eea
giving large regions of parameter space in which  light lepton events are dominant, particularly for an inverted hierarchy. 

A promising way to study the doubly-charged fermion at the LHC is via neutral-current pair-production,  giving 
\bea
pp\ \rightarrow\ \overline{\f^{--}}\f^{--}\ \rightarrow\ \ell^+W^+\ell^- W^-\ \rightarrow\  \ell^+\ell^+\nu \ell^-+\mathrm{jets},\label{f--_signal}
\eea
where one $W$ decays leptonically to enable charge identification and  the other decays hadronically to enable mass reconstruction. In optimistic cases with an inverted hierarchy the branching fraction for light charged-lepton events is $\sim1$ and the only significant branching-fraction suppression of the signal comes from  decaying the $W$'s. In the more pessimistic case of a normal hierarchy the branching fraction suppression can be important.

The results of Ref.~\cite{delgado_et_al_triplet} allow one to deduce that, for large regions of parameter space, there exists good discovery potential for $\f^{--}$ at the LHC.  Ref.~\cite{delgado_et_al_triplet}  identified the process \eqref{f--_signal} as a leading way to probe the doubly-charged fermion $\f_T^{--}\in\f_T\sim(1,3,-2)$, and their analysis appears to carry through for $\f^{--}$ in the present model, modulo three exceptions; the coupling between $\f_T$ and the $Z$ boson differs from $g_{\lz\lf--}$ by a factor of $g_{\lz\lf--}^2/g_{\lz T--}^2\simeq3.7$, increasing the neutral-current cross section for pair production in the present model;  Ref.~\cite{delgado_et_al_triplet} assumed equal branching fractions of $1/3$ for decays of $\f_T^{--}$ to distinct charged leptons; and there are additional background events in the present model from the process
\bea
pp\ \rightarrow\ \overline{\f^{0}}\f^{0}\ \rightarrow\ \ell^+W^-\ell^- W^+\ \rightarrow\  \ell^+\ell^-\nu \ell^-+\mathrm{jets}.\label{f0_signal_background}
\eea
However,  the production cross section for this background is always smaller than the signal cross section by an $\mathcal{O}(1)$ factor (see Figure~\ref{fig:pp_Z_production}). Furthermore the branching-fraction suppression when $\f$ decays is  more severe for the background process \eqref{f0_signal_background} than the signal \eqref{f--_signal} (by a factor of~$\lesssim 1/4$; see Tables~\ref{NH_light_charged_lepton_signals} and \ref{IH_light_charged_lepton_signals}). Selections should  further suppress the background;~Ref~\cite{delgado_et_al_triplet} imposes a mass cut requiring that the invariant mass $M_{\ell^-\ell^-\nu}$ be close to $M_{jj\ell^+}$, whereas the background from \eqref{f0_signal_background} gives $M_{\ell^-\ell^+\nu}$ close to $M_{jj\ell^+}$. Thus, the additional background should not significantly diminish ones ability to extract a signal from the process \eqref{f--_signal}.
%---------------------------------------------------------
%----------------------------------------------------

Given that the production cross section is larger in the present model,  one  expects the discovery reach  for $\f^{--}$ at the LHC to exceed the discovery reach for $\f_T^{--}$ quoted in Ref~\cite{delgado_et_al_triplet} for the entire region of parameter space with $BR(\f^{--}\rightarrow (e~\mathrm{or}~\mu)+W^-)\ge2/3$. This includes most of the parameter space for the case of an inverted hierarchy and a sizable fraction of the parameter space for a normal hierarchy, giving   a $5\sigma$ discovery reach  of at least $M_\f\lesssim 500$~(700)~TeV   with $10^2$~fb$^{-1}$ of data at the 7~(14)~TeV LHC for $BR(\f^{--}\rightarrow (e~\mathrm{or}~\mu)+W^-)=2/3$~\cite{delgado_et_al_triplet}.  Greater (reduced) reach is expected as one increases (decreases) this branching fraction and/or the integrated luminosity; see Figure~7 of Ref~\cite{delgado_et_al_triplet}. For example, in the most optimistic case of an inverted hierarchy with $BR(\f^{--}\rightarrow (e~\mathrm{or}~\mu)+W^-)\sim1$ the $5\sigma$ discovery reach will be at least $M_\f\lesssim 1$~TeV   with $10^2$~fb$^{-1}$ of data at the 14~TeV LHC.

These numbers are a rough guide but none the less suggest that good discovery reach can be achieved at the LHC.  It would be interesting to refine these numbers and undertake a detailed comparative analysis of  the exotic fermions in all the seesaw/radiative models.  Note that each of the models $(a)$ through $(d)$ in Table~\ref{table:seesaw_models} contain a doubly-charged fermion. Therefore an LHC  search  for exotic doubly-charged fermions via the process \eqref{f--_signal}  would give generic bounds applicable to all of the models (or, optimistically,  generate evidence for the origin of neutrino mass). Such an analysis appears to be the simplest experimental search capable of probing all of the models.

%%%%%%%%%%%%%%%%%%%%%%%%%%%%%%%%%%%%%%%%%%%
\section{Conclusion\label{sec:conc}}

The Type-I and Type-III seesaw mechanisms are part of a larger, generalized set of tree-level seesaws that can achieve small neutrino mass via heavy fermion exchange~\cite{McDonald:2013kca}. The generalized seesaw mechanisms differ from their simpler counterparts in two main ways; firstly, they realize mass via  low-energy effective operators with mass-dimension $d>5$, and, secondly, they  allow radiative  masses that can dominate the seesaw mass in  regions of parameter space. The models are therefore something of a hybrid between the traditional seesaw mechanisms and the traditional radiative models of neutrino mass; see Figure~\ref{fig:2venn}. 

There are only four minimal model of this type that give effective operators with $d\le9$~\cite{McDonald:2013kca}. In this work we studied the remaining $d=9$ model, detailing the origin of neutrino mass and investigating the collider phenomenology of the exotic quadruplet fermions predicted in the model [$\f\sim(1,4,-1)$]. Collider production of these exotics proceeds via electroweak interactions with cross sections  that are fixed, modulo their dependence on the fermion mass scale. The decay properties of the fermions encodes information about their connection to neutrino mass; for example the branching ratio for $\f^+\rightarrow \ell^+ Z$ is highly suppressed relative to $BR(\f^+\rightarrow \nu W^+)$ due to the suppression of $\langle S^0\rangle$ inherent in the model. Furthermore the branching fractions for charged leptons of different families are sensitive to both the neutrino mass hierarchy and the mixing phases. 

Lepton number violating like-sign dilepton events, which have no SM background, are not expected to be observable in the model.  This provides a powerful way to discriminate between the present model and the alternative tree-level seesaws with a heavy intermediate fermion (like the Type-III seesaw and models $(b)$ and $(d)$), which all predict like-sign dilepton events with observable branching fractions. This interesting difference reuslts from the VEV hierarchy inherent in the present model. Our analysis suggests that the doubly-charged exotic $\f^{--}$ could be discovered at the $5\sigma$ level for $M_\f\lesssim500$~(700)~GeV with 100~fb$^{-1}$ of data at the 7 (14)~TeV LHC. This reach extends higher in more optimistic  regions of parameter space. We briefly discussed ways to differentiate the distinct fermions in the various seesaw/radiative models, and noted that a dedicated LHC search for exotic doubly-charged fermions would provide generic bounds applicable to all of the models (or discover evidence for the origin of neutrino mass). It would be interesting to consider a more detailed comparative analysis of the exotic fermions in the four $d\le9$ models, to better illustrate both the LHC discovery reach and ways to discriminate the models.

Note added: After completion of this work Ref.~\cite{Ma:2013tda} appeared, in which the analysis of Ref.~\cite{delgado_et_al_triplet} was extended.
%%%%%%%%%%%%%%%%%%%%%%%%%%%%%%%%%%%%%%%%%%%%%%%5
\section*{Acknowledgments\label{sec:ackn}}
The author thanks R.~Foot, J.~Heeck, G.~Y.~Jeng, L.~Jong, A.~Kobakhidze, I.~Picek, A.~Saavedra, and W.~Skiba. This work was supported by the Australian Research Council.
%%%%%%%%%%%%%%%%%%%%%%%%%%%%%%%%%%%%%%%%%%%%%%%5
\appendix
%%%%%%%%%%%%%%%%%%%%%%%%%%%%%%%%%%%%%%%%%%%%%%%%%%%%%
\section{Expanded Lagrangian Terms\label{app:expanded_yukawa}}
The Yukawa Lagrangian for $\Delta$ can be expanded to obtain
\bea
& &\mathcal{L}\ \supset\ -\lambda_\Delta\,\bar{L} \f_R\Delta-\lambda_\Delta^\dagger \,\overline{\f_R}L\Delta\nonumber\\
&\equiv&-\lambda_\Delta\,\bar{L}^a\,(\f_R)_{abc}\,\epsilon^{cd}\,\Delta_d^{\ b}-\lambda_\Delta^\dagger\, (\overline{\f_R})^{abc}\,L_a\,\Delta_b^{\ d}\,\epsilon_{cd}\nonumber\\
&=&-\frac{\lambda_\Delta}{\sqrt{2}}\left\{ \Delta^- \left[\overline{\nu_L} \f_R^+ +\frac{1}{\sqth}\overline{e_L}\f^0_R\right]-\Delta^+ \left[\frac{1}{\sqth}\overline{\nu_L} F_R^- +\overline{e_L}\f^{--}_R\right]\right.+\left.\sqrt{\frac{2}{3}}\Delta^0 \left[\overline{\nu_L} \f_R^0 +\overline{e_L}F^{-}_R\right]\right\}\nonumber\\
&-&\frac{\lambda_\Delta^\dagger}{\sqrt{2}}\left\{ \Delta^+ \left[\overline{\f_R^+}\nu_L  +\frac{1}{\sqth}\overline{\f_R^0}e_L\right]-\Delta^- \left[\frac{1}{\sqth}\overline{F_R^-}\nu_L +\overline{\f_R^{--}}e_L\right]\right.+\left.\sqrt{\frac{2}{3}}\Delta^0 \left[\overline{\f_R^0}\nu_L +\overline{F^{-}_R}e_L\right]\right\},\nonumber\\
& &\label{delta_yukawa_lagrangian}
\eea
and similarly the Yukawa Lagrangian for $S$ gives
\bea
& &\mathcal{L}\ \supset\ -\lambda_{\s}\,\overline{L^c} \f_LS-\lambda_{\s}^\dagger\, \overline{\f_L}L^cS^*\nonumber\\
&\equiv&-\lambda_{\s}\,(\overline{L^c})^d\,S_{abcd}(\f_L)_{a'b'c'}\,\epsilon^{aa'}\,\epsilon^{bb'}\,\epsilon^{cc'}-\lambda_{\s}^*\, (\overline{\f_R})^{a'b'c'}\,(L^c)_d\,(S^*)^{abcd}\,\epsilon_{aa'}\,\epsilon_{bb'}\,\epsilon_{cc'}\nonumber\\
&=&-\lambda_{\s}\,\left\{\left[\overline{e_L^c}S^{+++} -\frac{1}{\sqrt{4}}\overline{\nu_L^c}S^{++}\right]\f_L^{--}-3 \left[\frac{1}{\sqrt{4}}\overline{e_L^c}S^{++} -\frac{1}{\sqrt{6}}\overline{\nu_L^c}S^{+}\right]\frac{1}{\sqrt{3}}F_L^{-}\right.\nonumber\\
& &\qquad+\left.3 \left[\frac{1}{\sqrt{6}}\overline{e_L^c}S^{+} -\frac{1}{\sqrt{4}}\overline{\nu_L^c}S^{0}\right]\frac{1}{\sqrt{3}}\f_L^{0}-\left[\frac{1}{\sqrt{4}}\overline{e_L^c}S^{0} -\overline{\nu_L^c}S^{-}\right]\f_L^{+}\right\}+\mathrm{H.c.}\nonumber
\eea
We suppress flavor labels in the above.
The quartic scalar coupling responsible for inducing a nonzero VEV for $S$ is expanded as follows:
\bea
& &V\supset -\lambda \tilde{H}^\dagger \Delta S^* H +\mathrm{H.c.}\nonumber\\
&\equiv& -\lambda\, (\tilde{H}^\dagger)^{a'}\, \Delta_b^{\ c'} \, (S^*)^{abcd}\, H_d\,\epsilon_{aa'}\, \epsilon_{cc'}+\mathrm{H.c.}\nonumber\\
&=&-\frac{\lambda}{2}H^+H^+ \left\{\frac{1}{\sqrt{3}}(S^+)^*\Delta^-+S^{--}\Delta^0-\sqrt{2}S^{---}\Delta^+\right\}\nonumber\\
& &\quad-\ \frac{\lambda}{\sqrt{2}}H^+H^0\left\{S^{0*}\Delta^-+\sqrt{\frac{2}{3}}(S^+)^*\Delta^0-S^{--}\Delta^+\right\}\nonumber\\
& &\quad -\ \frac{\lambda}{\sqrt{2}}H^0H^0\left\{(S^-)^*\Delta^-+\frac{1}{\sqrt{2}}S^{0*}\Delta^0-\frac{1}{\sqrt{6}}(S^+)^*\Delta^+\right\}+\mathrm{H.c.}
\eea
%%%%%%%%%%%%%%%%%%%%%%%%%%%%%%%%%%%%%%%%%%%%%%%%%%%%%
\section{Limits for the Loop Mass\label{sec:loop_mass_limits}}
Here we present approximate expressions for the loop mass in a number of limits. The general expression for the loop mass is
\bea
(\mathcal{M}^{loop}_\nu)_{\ell\ell'}&\simeq&\frac{(2-3\sqrt{2})}{6}\,\frac{[(\lambda^*_\Delta)_{\ell } \,(\lambda_{\s})_{\ell'} + (\lambda_{\s})_{\ell } \,(\lambda^*_\Delta)_{\ell'}]}{32\pi^2}\,M_{\f}\times\nonumber\\
& &\quad\quad\frac{\lambda\,\langle H^0\rangle^2}{M_{S}^2-M_{\Delta}^2}
\left[ \frac{M^2_{S}}{M_{\mathcal{F}}^2-M_S^2}\,\log \frac{M_{\mathcal{F}}^2}{M_{S}^2}\ -\ (M_S\rightarrow M_\Delta)\right].
\eea
Writing this mass as
\bea
(\mathcal{M}^{loop}_\nu)_{\ell\ell'}&\simeq&\frac{(2-3\sqrt{2})}{6}\,\frac{[(\lambda^*_\Delta)_{\ell } \,(\lambda_{\s})_{\ell'} + (\lambda_{\s})_{\ell } \,(\lambda^*_\Delta)_{\ell'}]}{32\pi^2}\,\lambda\,\langle H^0\rangle^2\times\mathcal{I},
\eea
we can determine the factor $\mathcal{I}$ for various  limits of interest. 
\bea
\mathcal{I}&\simeq& \frac{1}{2M_{\f}}\qquad\qquad\qquad\quad\qquad\mathrm{for}\quad M^2_{\f}\simeq M_\Delta^2\simeq M_{\s}^2,\nonumber\\
\mathcal{I}&\simeq&\frac{M_{\f}}{M_{\s}^2-M_\Delta^2}\,\log\frac{M_{\s}^2}{M_\Delta^2} \quad\qquad\mathrm{for}\quad M^2_{\f}\ll M_\Delta^2,\, M_{\s}^2\,,\nonumber\\
\mathcal{I}&\simeq&\frac{M_{\f}}{M_{\s}^2}\qquad\qquad\qquad\quad\qquad\mathrm{for}\quad M^2_{\f}\ll M_\Delta^2\approx M_{\s}^2\,,\nonumber\\
\mathcal{I}&\simeq&-\frac{M_{\f}}{M_{\Delta}^2}
\left[ \frac{M^2_{\s}}{M_{\mathcal{F}}^2-M_\s^2}\,\log \frac{M_{\mathcal{F}}^2}{M_{\s}^2}\ -\ \log \frac{M_{\mathcal{F}}^2}{M_{\Delta}^2}\right]\qquad\mathrm{for}\qquad M_{\f}^2\,,M_\s^2\ll M_\Delta^2.\nonumber\\
\eea

%%%%%%%%%%%%%%%%%%%%%%%%%%%%%%%%%%%%%%%%%%%%%%%%%%%%%

%%%%%%%%%%%%%%%%%%%%%%%%%%%%%%%%%%%%%%%%%%%
%%%%%%%%%%%%%%%%%%%%%%%%%%%%%%%%%%%%%%%%%%%%%%%%%%%

\end{document}